\documentclass[11pt]{article}

\usepackage[utf8]{inputenc}
\usepackage[T1]{fontenc}
\usepackage{amsmath,amssymb,amsthm}
\usepackage{natbib}
\usepackage{hyperref}
\usepackage[margin=1in]{geometry}
\usepackage{graphicx}
\usepackage{booktabs}



\newtheorem{proposition}{Proposition}
\newtheorem{theorem}{Theorem}
\newtheorem{lemma}{Lemma}
\newtheorem{corollary}{Corollary}
\theoremstyle{remark}
\newtheorem{remark}{Remark}

\title{Model-assisted estimation with a training subsample:\\
a two-phase sampling approach with design-based variance estimation}

\author{Mar\'ia Eugenia Ria\~no \\[2pt]
\small Department of Quantitative Methods, FCEA,\\
\small Universidad de la Rep\'ublica, Montevideo, Uruguay}

\date{}

\begin{document}

\maketitle

\begin{abstract}
\noindent When a flexible prediction model is fitted on a training
subsample drawn from a probability sample, the model-assisted
estimator actually reported arises from one realized partition, yet
existing theory quantifies uncertainty only for partition-averaged,
cross-fitted, or symmetrized versions of it. We represent the
training subsample as a second phase of sampling and derive, exactly
and for any algorithm, a two-term variance decomposition and the
variance family linking the single-partition estimator to its
Rao--Blackwellized average, whose design bias it shares. For
cell-mean rules under equal-probability designs the estimator is,
exactly, a convex combination of a poststratified plug-in at the
training-sample size and the exactly unbiased single-split
estimator of Sanguiao Sande and Zhang, which locates its bias
precisely. For tree-type predictors the second-phase variance is
computable in closed form when the cell structure is fixed or
$s$-measurable, and in that regime, for cell counts $H=o(n)$, its
share of total variance grows with the number of cells --- a
design-based mechanism distinct from residual shrinkage and
consistent with the variance underestimation documented for
flexible learners. We propose an analytic and a replication
variance estimator, neither altering the point estimate, and
evaluate them by simulation under simple random and stratified
designs, in the adaptive regime and in an $s$-measurable
configuration in which the theory is exact: in the populations
studied, budgeting the second phase recovers much of the coverage
lost by ignoring it, and the residual shortfall is attributable to
the bias of the point estimator rather than to its variance.
\end{abstract}


\section{Introduction}
\label{sec:intro}

Model-assisted estimation offers a principled way of using auxiliary
information in design-based inference for finite populations: an assisting
model produces predictions for all population units, and a
probability-weighted correction term protects the estimator against
misspecification of that model \citep{sarndal1992}. Over the past two
decades, the assisting models have moved well beyond linear regression, to
include local polynomial \citep{breidt2000} and model-calibrated
\citep{wu2001,montanari2005} estimators, the modern prediction
techniques reviewed by \citet{breidt2017}, regression trees
\citep{mcconville2019}, and random forests \citep{dagdoug2023},
among others. The appeal of this program is
that the design-based guarantees --- approximate design unbiasedness and
design consistency --- are retained under mild conditions whatever the
quality of the assisting model, while a good model can deliver substantial
efficiency gains. The guarantees for \emph{uncertainty quantification},
however, are more fragile. Valid variance estimation with
machine-learning predictors remains, as \citet{haziza2025} notes, a
central open question for survey statisticians, and it is on the variance
side that flexible learners create difficulties which the classical theory
was not designed to handle.

The difficulty can be stated with minimal notation. Let $U$ be a finite
population with target total $Y=\sum_{k\in U} y_k$, and let $s$ be a
probability sample drawn from $U$ under a design $p(\cdot)$ with inclusion
probabilities $\pi_k$. A model-assisted estimator has the generic form
\begin{equation}
\label{eq:generic-ma}
\hat{Y} \;=\; \sum_{k\in U}\hat{m}(\mathbf{x}_k)
\;+\; \sum_{k\in s}\frac{y_k-\hat{m}(\mathbf{x}_k)}{\pi_k},
\end{equation}
a census-level prediction component plus a probability-sample correction
based on prediction residuals. When the prediction rule $\hat{m}$ is
itself fitted on the same sample $s$ that enters the correction term, the
two components are no longer design-independent: every residual in the
correction is an \emph{in-sample} residual of a rule chosen, in part, to
make exactly those residuals small. The standard asymptotic argument
sidesteps this dependence by treating $\hat{m}$ as if it converged to a
fixed population-level rule, so that the fitted predictions are
effectively treated as fixed in the variance approximation. For stable,
low-complexity models this is innocuous. For flexible learners it is not:
a rule that adapts closely to $s$ produces in-sample residuals that
systematically understate the prediction errors on $U\setminus s$, and the
usual plug-in variance estimator inherits that understatement.
\citet[Sec.~6.2]{dagdoug2023} document this phenomenon for
random-forest-assisted estimation and propose, as a remedy, a variance
estimator based on $K$-fold (out-of-fold) residuals, while noting
explicitly that establishing the theoretical properties of that estimator
requires further research. Independently, \citet{cosenza2025} show
empirically, in model-assisted forest inventory with several learners,
that overfit internally-fitted models lead to both bias and variance
underestimation, with resampling-based variance estimators only partially
robust. Neither reference suggests that the point estimator collapses;
the failure they document concerns the variance statement that accompanies
it --- which is precisely what design-based inference is supposed to
certify.

A natural response to this diagnosis is to break the dependence by
construction: fit the prediction rule on a \emph{training subsample}
$s_1\subset s$, selected from $s$ by a known randomization mechanism
$q(\cdot\mid s)$, and let the correction term rely on residuals whose
out-of-sample character can be controlled. Doing so, however, introduces
a second layer of randomness --- the estimator now depends on two nested
random selections, $s$ from $U$ under $p$ and $s_1$ from $s$ under $q$
--- and it is in the treatment of this second layer that the recent
literature divides into three families of solutions.

The first family removes the second layer by \emph{averaging over it}.
\citet{sanguiao2021} introduce a subsampling Rao--Blackwell method
delivering exactly design-unbiased estimation with linear or nonlinear
learners: the estimator based on a single realized subsample is replaced
by its conditional expectation over $q$ given $s$, approximated in
practice by averaging over $B$ independent subsample draws and refits.
\citet{zhang2026} develop this program into a general ``$pq$-design''
framework --- $s$ drawn from $U$ under $p$, $s_1$ drawn from $s$ under $q$
--- with conditions for representative training, subsampling
Rao--Blackwellized predictors, and design-unbiased estimators of mean
squared error based on out-of-bag ideas. The Rao--Blackwell step is, by
design, a variance-reduction device: it eliminates the $q$-variability of
the fitted rule by integration. The resulting inferential object is
therefore the \emph{partition-averaged} estimator; the estimator produced
by one realized split appears only as an intermediate quantity whose
randomness the method is constructed to remove, at the computational price
of $B$ refits of the learner. Interval estimation with design-based
coverage under this framework is, by these authors' own account, still
open.

The second family neutralizes the dependence \emph{asymptotically} through
cross-fitting, importing the sample-splitting logic of
double/debiased machine learning \citep{chernozhukov2018} into
survey estimation; on the randomness of the split itself in that
literature, and the practice of aggregating over splits, see
\citet{rinaldo2019}.
\citet{dagdoug2026} formalize the cross-fitted model-assisted estimator
--- the population is partitioned at random into $K$ folds before
sampling, and the residual of each sampled unit is computed from a
prediction rule fitted on the sampled units of the other folds ---
together with a variance estimator that replaces census residuals
by cross-fitted residuals, within a Neyman-orthogonality framework adapted
to finite populations. \citet{kwon2026} provide, for linear regression, a
complete theory of this construction (SREG): first-order equivalence with
an oracle difference estimator, asymptotic normality, and a consistent
variance estimator based on out-of-fold residuals, verified for simple
random, stratified, and rejective sampling. In a related direction, the
$K$-fold variance-estimation program announced by
\citet[Sec.~6.2]{dagdoug2023} has since been developed theoretically in
the context of imputation for item nonresponse
\citep{dagdoug2025}, though not for model-assisted estimation. Two
features of the cross-fitting family matter for our purposes. Every unit
plays both roles --- training in some folds, correction in others --- so
the construction deliberately combines $K$ complementary splits; and
the fold assignment is made at random and then held fixed, its effect
confined to a remainder shown to be asymptotically negligible
(\citealp[Sec.~3.3]{dagdoug2026}; \citealp[Theorems~4--5]{kwon2026}). What is characterized is the cross-fitted
estimator as a whole, not the estimator that results from a single
realized partition of the sample into a training part and a correction
part.

The third family absorbs the fitting uncertainty into an \emph{exact
variance computation}. \citet{dharamshi2025} observe that when the fitted
predictor admits a U- or V-statistic representation --- from linear rules
through certain ensembles --- the model-assisted estimator can itself be
represented as a U/V-statistic, whose design-based variance can then be
computed exactly, capturing the model-fitting uncertainty that plug-in
approximations ignore. The construction is developed in detail for the
linear GREG under Poisson sampling, and the authors identify extensions
beyond Poisson designs, as well as computationally efficient algorithms
for ensembles with external randomness, as open problems. The inferential
object here is again related to, but distinct from, ours: the U/V
representation symmetrizes the fitting over subsamples within the
variance calculation, rather than analyzing a procedure in which a
training subsample is drawn once, by an explicit and known mechanism $q$,
and the resulting single fitted rule is used.

These three strategies are not interchangeable, because they do not study
the same estimator. It is worth being explicit about the four objects in
play: (i) the estimator built from \emph{one realized partition} $(s,
s_1)$; (ii) its partition-averaged version, the expectation over the
distribution of $s_1$ given $s$; (iii) the cross-fitted estimator, which combines $K$
complementary folds so that no single split is privileged; and (iv) a
symmetrized U/V-statistic representation used for exact variance
calculation. The existing literature provides design-based theory for
(ii), (iii) and, in the linear Poisson case, (iv). Yet an analyst who
implements subsample-training \emph{once} --- as any production
implementation with an expensive learner naturally would --- obtains one
realized $s_1$, one fitted rule, and one estimate: object (i). To our
knowledge, no existing result provides the design-based variance of that
estimator under the joint $(p,q)$ randomization, nor a variance estimator
computable from the single realization actually observed. Existing
methods instead target the averaged, combined, or symmetrized objects.
This leaves unresolved the uncertainty statement that should accompany the
estimate a practitioner actually reports.

The estimator studied here is therefore not new as a formula ---
and this is the point. In its computing form (Lemma~\ref{lem:collapse}
below) it is the standard model-assisted estimator with the working
model fitted on a subsample of $s$: what an analyst obtains by
holding out part of the sample to train the learner. What the
two-phase representation adds is the variance that goes with that
familiar formula, and it makes visible a source of uncertainty that
the full-sample fit structurally hides: for cell-mean rules the
correction term of the full-sample estimator vanishes identically
(Section~\ref{sec:composite}), so the full-fit estimate rests
entirely on the fitted structure, and its dependence on that
particular fit never reaches the reported variance. Training on
$s_1$ restores a correction built from residuals whose
out-of-sample character is controlled, and turns the instability of
the algorithm into an explicit variance component, $E_p(V_2)$, that
can be estimated without conditions on the learner.

The question is not peculiar to the estimator studied here. Whenever
a sample is split and the reported quantity depends on which units
fell in the training part, the randomness of that split belongs in
its variance; this covers the single-partition estimator, and it
reaches cross-fitting through the fold assignment. It also enters
through model selection and not only through fitting: when the
held-out part is used to tune, select variables or decide when to
stop, the selected model is a function of the realized split, and
refitting afterwards on the full sample does not undo that
dependence. We do not treat the selection case here --- the rule
$\hat{m}_1$ studied below is a deterministic function of $s_1$ ---
but it is the same randomization entering by a second route.

The two-phase sampling framework of \citet[Chap.~9]{sarndal1992} supplies,
in principle, exactly the missing accounting device: viewing $s$ as a
first-phase and $s_1$ as a second-phase sample decomposes the variance
as $V_1 + \mathrm{E}(V_2)$, attributing to each randomization its own
share of the uncertainty. But the classical theory does not cover this
configuration: when the auxiliary information is common to both phases
--- the book's Case~1, exactly the subsample-training setting --- its
intermediate term vanishes identically, whereas here the two fits
differ through the \emph{fitting sample} rather than the information,
so the term survives and the classical results must be rederived.
Section~\ref{sec:setup-classical} develops the point, including the
reversal that makes a workable theory possible: quantities
incomputable in the classical setting, where $y$ is observed only at
the second phase, become computable here.

Three considerations make the single partition the natural object of
study. First, \emph{correspondence between estimate and inference}:
partition averaging changes the estimator, so a variance statement for
the averaged estimator does not refer to the number an analyst obtains
from one realized split. If inference is to refer to the estimate
actually produced, the second-phase randomization must be budgeted in
the variance, not integrated out of the estimator. Second,
\emph{production practice}: replication is routine in official
statistics for variance estimation \citep{wolter2007}, but a published
point estimate is ordinarily expected to trace back to one documented
procedure --- one fitted model, whose cells can be reported and
audited, a concern made explicit in recent quality frameworks for
statistical algorithms \citep{yung2022}. The single partition serves
that expectation; it is noisier than an average over partitions, and
its price is exactly $E_p(V_2)$ --- this paper shows how to measure
that price and carry it into the reported interval. Third,
\emph{computation}: the single partition requires one fit of the
learner; the analytic variance estimator below adds none when the
cell structure is fixed or $s$-measurable, and the replication
estimator adds $A$ refits of the second phase. How these budgets
compare with the $K$ fits of cross-fitting or the $B$ fits of
partition averaging at matched fit counts, and with the coverage each
buys, is a comparison we leave to a separate study. A variance
decomposition that keeps the second-phase component explicit also
answers a question the other approaches cannot pose: how much
precision does a single partition actually cost, and how does that
cost vary with the training fraction $n_1/n$?

This paper makes three contributions. First, we give a formal two-phase
representation of the training-subsample mechanism and derive the
resulting model-assisted estimator, denoted $\hat{Y}_{TS}$, whose
intermediate term --- a design-weighted contrast between the predictions
of the full-sample and training-sample fits --- does not vanish, unlike
its classical two-phase analogue; we characterize its design bias
exactly --- the single-partition estimator inherits the bias of its
partition average, which reduces the bias question to the one already
open for partition-averaged estimators --- and we obtain two exact
structural results: a closed-form reduction in a degenerate case
(prediction rule equal to the subsample mean) that isolates the
bias--variance trade-off induced by the partition, and, for cell-mean
rules under equal-probability designs, a representation of the
estimator as a convex combination of a poststratified plug-in and the
exactly unbiased single-split estimator of \citet{sanguiao2021},
which locates its bias in closed form. Second, we derive ---
rather than inherit from the classical two-phase formulas --- an
approximate design-based variance for $\hat{Y}_{TS}$ with an explicit
two-term decomposition into a first-phase component and the expectation
of a second-phase component, with conditions that are verifiable for
regression trees by exploiting their poststratification structure
\citep{mcconville2019}; and we propose two companion variance estimators:
an analytic estimator based on the decomposition, available in closed
form for tree-type predictors, and a replication estimator targeting the
second phase; neither alters the reported single-partition point
estimate. The decomposition itself is exact; what rests on the usual
model-assisted approximations is the evaluation of its first-phase
component. Third, we report a simulation study, on
populations comparable to those of \citet{dagdoug2023}, that evaluates the
proposed estimator and its variance estimators against the
Horvitz--Thompson, GREG, tree-assisted, single-split, Monte Carlo
Rao--Blackwellized, and $K$-fold cross-fitted alternatives ---
quantifying the efficiency cost of the single partition as a function
of the training fraction and the coverage of the resulting intervals,
under simple random and stratified first-phase designs, in the
adaptive configuration and in an $s$-measurable configuration in
which the tree-type theory applies exactly. Random forests appear
only as an empirical remark; their formal theory is beyond our scope
here.

These results divide along one line. The exact statements require no
conditions: Theorem~\ref{thm:decomp} decomposes the variance for any
algorithm, any design and any number of refits, and
Proposition~\ref{prop:rep-unbiased} estimates its second-phase
component without bias for any algorithm and any $A\geq 2$. The
asymptotic statements require high-level conditions --- (A4$'$) for
the bias, an $s$-measurable cell structure for the closed-form
variance --- and both remain open for adaptively fitted learners.

Although developed in the survey-sampling tradition, the question is
not confined to it: prediction-powered inference
\citep{angelopoulos2023} is, in sampling terms, the difference
estimator of \citet{cassel1976}, its power-tuned variant corresponds
to the GREG \citep{mozer2026}, a design-based recasting for spatial
populations treats the prediction map as fixed \citep{shirota2026},
and when such a pipeline trains its predictor on data that overlap
the labeled correction set and is run once rather than averaged over
many splits, the dependence studied here arises in the same form.

The remainder of the paper is organized as follows. Section~\ref{sec:setup}
introduces the notation and the two-phase representation of the
training-subsample mechanism, and makes precise both the connection to,
and the departure from, the classical two-phase theory.
Section~\ref{sec:estimator} derives the proposed estimator, interprets its
three components, presents the closed-form degenerate case, and relates
the estimator formally to its cross-fitted and partition-averaged
counterparts. Section~\ref{sec:properties} establishes the design-based
properties: bias and the approximate variance with its two-term
decomposition, with explicit conditions for regression trees.
Section~\ref{sec:varest} develops variance estimation, including
the resampling estimator.
Section~\ref{sec:simulation} reports the simulation study, and
Section~\ref{sec:discussion} concludes with limitations and extensions.
Proofs are collected in the Appendix, which opens with a short list
of the elementary facts they use repeatedly. R code reproducing all
numerical results is provided as supplementary material, together
with scripts that verify every exact identity of
Sections~\ref{sec:estimator}--\ref{sec:varest} and of the Appendix
by Monte Carlo simulation or, where the design is small enough, by
complete enumeration.

\section{Setup and two-phase representation of the training subsample}
\label{sec:setup}

\subsection{Population, designs, and probabilistic structure}
\label{sec:setup-structure}

Let $U=\{1,\ldots,N\}$ be a finite population. Associated with unit $k$
are a study value $y_k$ and a vector of auxiliary variables
$\mathbf{x}_k$, the latter known for every $k\in U$ (frame-level
auxiliary information). The target is the population total
$Y=\sum_{k\in U}y_k$.

Randomness enters in two nested stages, both under the analyst's
control or knowledge.

\emph{First phase.} A probability sample $s\subset U$ of fixed size $n$
is drawn under a sampling design $p(\cdot)$, with first- and
second-order inclusion probabilities $\pi_k=\Pr(k\in s)>0$ and
$\pi_{k\ell}=\Pr(k,\ell\in s)$, and sample membership indicators
$I_k=\mathbf{1}(k\in s)$. Write $\Delta_{k\ell}=\pi_{k\ell}-\pi_k\pi_\ell$,
$\hat{Y}_\pi=\sum_{k\in s}y_k/\pi_k$ for the Horvitz--Thompson (HT)
estimator \citep{horvitz1952}, and
$\hat{N}_\pi=\sum_{k\in s}1/\pi_k$ for the HT estimator of the
population size. The values $y_k$ are observed for all $k\in s$.

\emph{Second phase.} Given $s$, a training subsample $s_1\subset s$ of
size $n_1$ is drawn under a known randomization mechanism
$q(\cdot\mid s)$ chosen by the analyst, and $s_2=s\setminus s_1$
denotes the complementary (assessment) part of the sample. Our
working choice, maintained throughout unless stated otherwise, is
simple random sampling without replacement of fixed size $n_1$ from
$s$, denoted $q=\mathrm{SRSWOR}(n_1)$. We write
\begin{equation}
\label{eq:f1}
f_1=n_1/n\in(0,1)
\end{equation}
for the \emph{training fraction}: the fraction of the \emph{sample}
used for fitting.

The subscript in \eqref{eq:f1} is deliberate. In survey sampling the
unadorned symbol $f$ conventionally denotes the first-phase sampling
fraction $n/N$, an entirely different quantity, which in the
asymptotic framework of Section~\ref{sec:setup-assumptions} is
denoted $\kappa$. The two fractions are unrelated in origin and in
role: $f_1$ is chosen by the analyst and, as
Section~\ref{sec:properties} shows, governs the second-phase
variance, whereas $n/N$ is a feature of the survey design and enters
only through first-phase finite-population corrections. Neither
constrains the other, and both may appear in the same expression.

Expectation and variance under the first phase are denoted $E_p$ and
$V_p$; expectation and variance under the second phase, conditionally
on $s$, are denoted $E_q(\cdot\mid s)$ and $V_q(\cdot\mid s)$; and
$E=E_pE_q$ denotes total (design) expectation. All statements in this
paper are design-based: no model is assumed for the $y_k$, which are
treated as fixed constants. For any statistic $\hat{\theta}$ depending
on $(s,s_1)$, the law of total variance gives the
decomposition
\begin{equation}
\label{eq:tp-decomp}
V(\hat{\theta})
= V_p\!\left\{E_q(\hat{\theta}\mid s)\right\}
+ E_p\!\left\{V_q(\hat{\theta}\mid s)\right\}
\;\equiv\; V_1 + E_p(V_2),
\end{equation}
where $V_2=V_q(\hat{\theta}\mid s)$ is the (conditional) second-phase
variance. Decomposition \eqref{eq:tp-decomp} is the accounting device
announced in Section~\ref{sec:intro}: it attributes to each
randomization its own share of the uncertainty. The probabilistic
structure $(p,q)$ is the same as in classical two-phase sampling
\citep[Chap.~9]{sarndal1992} and coincides with the ``$pq$-design'' of
\citet{sanguiao2021} and \citet{zhang2026}; what differs across these
strands, and from the present paper, is the estimator built on top of
it and the component of \eqref{eq:tp-decomp} that each approach
targets.

\subsection{Prediction algorithms}
\label{sec:setup-algorithms}

A prediction algorithm $\mathcal{A}$ is a map taking a fitting dataset
$\{(y_k,\mathbf{x}_k):k\in d\}$, $d\subset U$, to a prediction function
$\hat{m}_d:\mathbf{x}\mapsto\hat{m}_d(\mathbf{x})$. Two fits play a
role in what follows:
\[
\hat{m}=\hat{m}_s
\quad\text{(full-sample fit)},
\qquad
\hat{m}_1=\hat{m}_{s_1}
\quad\text{(training fit)}.
\]
The full-sample fit is a function of $s$ only; the training fit depends
on $(s,s_1)$ and is the source of all second-phase randomness. We also
define the partition-averaged (Rao--Blackwellized) prediction rule
\begin{equation}
\label{eq:rb-rule}
\bar{m}_s(\mathbf{x}) = E_q\{\hat{m}_1(\mathbf{x})\mid s\},
\end{equation}
which is $s$-measurable.

We use the term $s$-measurable throughout in its standard sense, and
state it once explicitly because it carries most of the bookkeeping
in what follows: a quantity is \emph{$s$-measurable} when it is a
function of the realized first-phase sample $s$ alone --- formally,
measurable with respect to the $\sigma$-field generated by $s$ ---
so that it is random before $s$ is drawn but \emph{fixed once $s$ is
known}, and in particular does not depend on the second-phase draw
$s_1$. Operationally this is the property that matters: an
$s$-measurable quantity passes through $E_q(\cdot\mid s)$ and
$V_q(\cdot\mid s)$ as a constant. Thus $\hat{m}$, $\hat{Y}_\pi$,
$\hat{N}_\pi$, the inclusion indicators $I_k$ and the rule
$\bar{m}_s$ of \eqref{eq:rb-rule} are all $s$-measurable, while
$\hat{m}_1$ is not. For tree-type algorithms --- our theoretical
working case --- the fitted rule is a poststratification: a regression
tree partitions the covariate space into terminal nodes and predicts,
within each node, an average of the fitted $y$-values
\citep{mcconville2019}. Predictions of such rules lie in the convex
hull of the fitted $y$-values, so boundedness of $y$ implies
boundedness of all fitted rules; random forests average many such
trees and inherit the same property.

Throughout, the study variable is continuous and the algorithm
returns a numerical prediction. Classification raises separate
issues --- a class count is the total of an indicator, so the
estimand carries over, but majority-vote rules are not cell means,
and a $q$ stratified by class label changes the law of the cell
counts $n_{1h}$ --- and is not treated here.

\subsection{Relation to classical two-phase sampling}
\label{sec:setup-classical}

Although \eqref{eq:tp-decomp} is the classical two-phase identity, the
present setting inverts the classical data structure in three respects,
and the inversion matters for which results transfer and which do not.
The summary is that the two-phase template returns a familiar
estimator but not a familiar variance: collapsing it yields the
standard model-assisted form fitted on $s_1$
(Lemma~\ref{lem:collapse}), while the intermediate term that the
classical theory eliminates survives here and carries a variance
component of its own.

First, in classical two-phase sampling the study variable is observed
only at the second phase; here $y_k$ is observed on \emph{all} of $s$,
and the second phase is an analyst's device rather than a data
constraint. The asymmetry has a practical consequence, anticipated in
Section~\ref{sec:intro}: quantities that are incomputable in the
classical setting become computable here. The census-level fit on
$s$ is the counterpart of the coefficient $\hat{\mathbf{B}}_{1s_a}$
of \citet[eq.~9.7.12]{sarndal1992}, which the classical theory
cannot compute because the $y_k$ are known only in the second-phase
sample \citep[p.~361]{sarndal1992}, and for which
\citet[Remark~9.7.2]{sarndal1992} offer only an alternative
estimator.

Second, the auxiliary information is \emph{identical} at both phases:
both fits use the same frame-level $\mathbf{x}$. In the classical
theory the two prediction stages are fitted on the same second-phase
sample and differ through nested auxiliary information, richer at
the second phase, and the regression estimator
\citep[eq.~9.7.2]{sarndal1992} contains, between the census term and
the residual correction, the intermediate term
\[
\sum_{k\in s}
\frac{\hat{m}^{(2)}(\mathbf{x}_k)-\hat{m}^{(1)}(\mathbf{x}_k)}{\pi_k},
\]
written in our notation, with $\hat{m}^{(2)}$ and $\hat{m}^{(1)}$ the
predictions from the richer and the common auxiliary vectors. When
the information is common to both phases and the two fits share
their variance specification --- Case~1 of
\citet[Sec.~9.7]{sarndal1992} --- the two predictions coincide and
the term vanishes identically \citep[eq.~9.7.24]{sarndal1992}. Here
the auxiliary information is likewise common, but the two
predictions differ through the \emph{fitting sample} ---
$\hat{m}=\hat{m}_s$ against $\hat{m}_1=\hat{m}_{s_1}$ --- an axis
absent from the classical theory, so the contrast is not identically
zero and the term survives. Regrouped with the census term taken
from the full-sample fit, it becomes the term $T_1$ of
Section~\ref{sec:estimator}; with common auxiliary information, the
classical theory has no nonvanishing counterpart for it.

Third, the second-phase design $q$ is entirely known and chosen by the
analyst, so second-phase moments can be computed --- and in favorable
cases evaluated exactly --- rather than estimated
(Section~\ref{sec:varest}). The classical two-phase
variance-estimation literature --- \citet{hidiroglou1998} on the
use of auxiliary information, \citet{fuller1998} and \citet{kim2006}
on replication methods, \citet{beaumont2015} on the conditional
decompositions --- operates on the classical information structure;
the reversal above is why its results are rederived here rather than
imported.

Because of the first two points, the results of
\citet[Result~9.7.1]{sarndal1992} on approximate unbiasedness and
approximate variance of the two-phase regression estimator do not
apply to the estimator studied here; Sections~\ref{sec:estimator}
and~\ref{sec:properties} rederive the corresponding properties from
first principles.

\subsection{Asymptotic framework and assumptions}
\label{sec:setup-assumptions}

Exact statements below (Lemma~\ref{lem:collapse},
Propositions~\ref{prop:degenerate}, \ref{prop:rb-identity},
\ref{prop:bias}(a), \ref{prop:v2-known}, \ref{prop:rep-unbiased}, and
Theorems~\ref{thm:decomp}, \ref{thm:trees}) require no asymptotic
framework. For approximation results we adopt the standard
finite-population asymptotic framework \citep{isaki1982}: a sequence of
populations and designs indexed by $\nu$, with
$N=N_\nu\to\infty$, $n=n_\nu\to\infty$, and $n/N\to\kappa\in[0,1)$;
the index $\nu$ is suppressed. The following conditions are used where
indicated.

\medskip
\noindent\textbf{(A1)} \emph{(Design regularity.)} There are constants
$\lambda>0$ and $C_1<\infty$ such that $\pi_k\geq\lambda\, n/N$ for all
$k\in U$ and
$\max_{k\neq \ell}|\Delta_{k\ell}|/(\pi_k\pi_\ell)\leq C_1/n$.

\medskip
\noindent\textbf{(A2)} \emph{(Boundedness.)} There is $C_y<\infty$ with
$\max_{k\in U}|y_k|\leq C_y$, and $\mathcal{A}$ returns predictions in
the convex hull of the fitted $y$-values (satisfied by trees and
forests).

\medskip
\noindent\textbf{(A3)} \emph{(Second-phase design.)}
$q=\mathrm{SRSWOR}(n_1)$ with $n_1=\lceil f_1 n\rceil$ for a target
training fraction $f_1\in(0,1)$ held fixed along the
sequence.\footnote{Because $n_1$ is an integer, the realized fraction
satisfies $f_1\leq n_1/n<f_1+1/n$, so the compact ratio
$(1-f_1)/f_1$ used below is exact only when $f_1n\in\mathbb{N}$; in
general
\begin{equation}
\label{eq:rounding}
1/n_1-1/n\;\leq\;(1-f_1)/(f_1 n),
\end{equation}
with equality if and only if $f_1n\in\mathbb{N}$. Exact
finite-sample statements are accordingly written in terms of $n_1$
(or $n_{1h}$ by cell), and $(1-f_1)/f_1$ is used where an asymptotic
order or an explicit bound is intended. The gap is $O(n^{-2})$ and
is negligible except when cells are small, a case revisited in
Corollary~\ref{cor:relative-cost}.}

\medskip
\noindent\textbf{(A4)} \emph{(Algorithm stability.)} There is a
deterministic sequence of bounded functions $m_\nu$ such that
\[
\delta_n^2
:= E\!\left[\frac{1}{N}\sum_{k\in U}
\{\bar{m}_s(\mathbf{x}_k)-m_\nu(\mathbf{x}_k)\}^2\right]
\longrightarrow 0 .
\]

\medskip
\noindent\textbf{(A4$'$)} \emph{(First-order equivalence.)} With
$\hat{Y}_{\mathrm{dif}}=\sum_{k\in U}m_\nu(\mathbf{x}_k)
+\sum_{k\in s}\{y_k-m_\nu(\mathbf{x}_k)\}/\pi_k$ the (infeasible)
difference estimator built from the limit rule,
$E[\{N^{-1}(\hat{Y}_{RB}-\hat{Y}_{\mathrm{dif}})\}^2]=o(n^{-1})$, where
$\hat{Y}_{RB}$ is defined in \eqref{eq:rb-estimator} below.

\medskip
Conditions (A1)--(A2) are standard \citep{breidt2017,dagdoug2023}.
Conditions (A4)--(A4$'$) are high-level stability conditions of the
type used throughout the model-assisted literature
\citep{breidt2017,sanguiao2021,dagdoug2023}; they restrict the
algorithm, not the $y$-values. We do not attempt to verify (A4$'$) for
adaptively grown trees or forests --- precisely the frontier discussed
in Section~\ref{sec:intro} --- but Section~\ref{sec:properties} verifies
the second-phase results, which are the paper's contribution, exactly
or with explicit remainders for tree-type predictors, without invoking
(A4$'$).

\subsection{Conditional and unconditional inference}
\label{sec:conditional}

The variance target adopted throughout is unconditional over both
phases: $V_1+E_p(V_2)$ in \eqref{eq:tp-decomp}. Since the analyst
chooses to randomize, records the seed, and could argue for
inference conditional on the realized split, the choice deserves an
argument rather than an assumption --- the more so since
conditional analyses of poststratification have a distinguished
tradition \citep{holt1979,rao1985}. Conditionally on $s_1$ the
fitted rule is fixed, and a conditional analysis is available in
principle; it is, in effect, what a fixed-prediction analysis
performs. But the split is a poor conditioning statistic here. It
is not ancillary: marginally $s_1$ is itself a probability sample
from $U$ (fact~(P4) of the Appendix), and the realized split
determines the fitted rule --- in the conditional-inference idiom,
it is informative about the realized precision only through the
fitted cells themselves. A conditional variance statement certifies
the dispersion that would arise if the same units were re-designated
as training data, a replication no production system performs;
replications of the documented procedure draw $q$ afresh, and the
variance that describes them is $V_1+E_p(V_2)$. The practical
question that motivates conditioning --- how much precision did this
one realized partition cost? --- is Corollary~\ref{cor:cost}'s, and
it has an unconditional answer. Integrating over $q$ is what turns
the instability of the fitted rule into a reportable variance
component rather than a hidden one.

\section{The proposed estimator}
\label{sec:estimator}

\subsection{Construction and basic structure}
\label{sec:estimator-construction}

Applying the two-phase regression template of
\citet[Sec.~9.7]{sarndal1992} to the structure of
Section~\ref{sec:setup} --- census-level predictions from one fit, an
intermediate correction contrasting the two fits, and a
probability-weighted residual correction --- yields the proposed
estimator
\begin{equation}
\label{eq:ysl-threeterm}
\hat{Y}_{TS}
= \underbrace{\sum_{k\in U}\hat{m}(\mathbf{x}_k)}_{T_0}
\;+\;
\underbrace{\sum_{k\in U}\Bigl(1-\frac{I_k}{\pi_k}\Bigr)
\{\hat{m}_1(\mathbf{x}_k)-\hat{m}(\mathbf{x}_k)\}}_{T_1}
\;+\;
\underbrace{\sum_{k\in s}\frac{y_k-\hat{m}(\mathbf{x}_k)}{\pi_k}}_{T_2}.
\end{equation}
The three components have distinct roles. $T_0$ is the census
prediction total under the full-sample fit. $T_2$ is the usual
probability-sample correction of the full-sample fit, so that
$T_0+T_2$ is exactly the generic model-assisted estimator of
Section~\ref{sec:intro} --- the estimator whose variance is
underestimated when the fitted predictions are treated as fixed.
$T_1$ is the intermediate term: a design-weighted contrast between the
training fit and the full-sample fit, with weights
$1-I_k/\pi_k$ whose first-phase expectation is zero for any fixed
function. It is, up to regrouping, the intermediate term that vanishes
identically in the classical Case~1 configuration
(Section~\ref{sec:setup-classical}) and survives here because the
two predictions differ through the fitting sample rather than
through the auxiliary information.

The representation \eqref{eq:ysl-threeterm} is an analytical device
rather than a computing formula, as the following identity shows.

\begin{lemma}
\label{lem:collapse}
For any $s$-measurable prediction rule in place of $\hat{m}$
(in particular the full-sample fit),
\begin{equation}
\label{eq:ysl-collapsed}
\hat{Y}_{TS}
= \sum_{k\in U}\hat{m}_1(\mathbf{x}_k)
+ \sum_{k\in s}\frac{y_k-\hat{m}_1(\mathbf{x}_k)}{\pi_k}.
\end{equation}
Consequently, $\hat{Y}_{TS}=(T_0+T_2)+T_1$, where $T_0$ and $T_2$ are
$s$-measurable and all second-phase randomness is carried by $T_1$.
\end{lemma}

Lemma~\ref{lem:collapse} admits three readings that organize the
rest of the paper. First, the collapsed form \eqref{eq:ysl-collapsed} is exactly
the generic model-assisted estimator with the training fit in place of
the full fit: predictions everywhere from $\hat{m}_1$, residual
correction over all observed units. This is the estimator a
practitioner naturally computes after fitting on a training subsample;
nothing about it is exotic, and it uses every observed $y_k$.
Second, $\hat{Y}_{TS}=(T_0+T_2)+T_1$ exhibits the proposed estimator
as the \emph{naive} full-fit estimator plus a second-phase correction
whose conditional variance $V_q(T_1\mid s)$ is precisely the component
that a fixed-prediction analysis omits. Third, because
\eqref{eq:ysl-threeterm} holds for any $s$-measurable reference rule,
the full-sample fit $\hat{m}$ never needs to be computed: it serves
only to interpret the decomposition.

\subsection{The degenerate case}
\label{sec:estimator-degenerate}

The simplest algorithm --- return the mean of the fitted $y$-values,
$\hat{m}_d(\mathbf{x})\equiv\bar{y}_d$ --- admits a closed form that
anchors the interpretation of everything that follows. Two familiar
procedures produce it. A regression tree grown with no splits ---
\texttt{rpart} with $\mathit{cp}=1$, or a minimum split size
exceeding $n_1$ --- predicts $\bar{y}_{s_1}$ at every $\mathbf{x}$;
so does least squares under an intercept-only working model
$y=\mu+\varepsilon$, whose fitted coefficient is
$\hat{\mu}=\bar{y}_{s_1}$. Neither is a recommendation: both are
limiting members of the families considered below, included because
they isolate the mechanism.

\begin{proposition}
\label{prop:degenerate}
Let $\mathcal{A}$ return the fitted-set mean, so that
$\hat{m}_1(\mathbf{x})\equiv\bar{y}_{s_1}$. Then, exactly,
\begin{equation}
\label{eq:degenerate}
\hat{Y}_{TS} = \hat{Y}_\pi + (N-\hat{N}_\pi)\,\bar{y}_{s_1}.
\end{equation}
In particular: (a) under simple random sampling at the first phase,
$\hat{N}_\pi=N$ and $\hat{Y}_{TS}=\hat{Y}_\pi$ exactly, so the
partition has no effect whatsoever; (b) under (A3),
$E_q(\hat{Y}_{TS}\mid s)=\hat{Y}_\pi+(N-\hat{N}_\pi)\bar{y}_s$, so the
second phase contributes variance but no conditional bias; and (c)
under (A3),
\begin{equation}
\label{eq:degenerate-v2}
V_q(\hat{Y}_{TS}\mid s)
= (N-\hat{N}_\pi)^2\left(\frac{1}{n_1}-\frac{1}{n}\right)S^2_{y,s},
\qquad
S^2_{y,s}=\frac{1}{n-1}\sum_{k\in s}(y_k-\bar{y}_s)^2,
\end{equation}
exactly, and consequently, by \eqref{eq:rounding},
\begin{equation}
\label{eq:degenerate-v2-f}
V_q(\hat{Y}_{TS}\mid s)
\;\leq\;
\frac{1-f_1}{f_1}\cdot\frac{(N-\hat{N}_\pi)^2 S^2_{y,s}}{n},
\end{equation}
with equality when $f_1 n\in\mathbb{N}$.
\end{proposition}

Proposition~\ref{prop:degenerate} isolates, in closed form, how the
training partition enters: through the product of the level error
$N-\hat{N}_\pi$ of the first phase and the subsampling noise of the
training mean. For maximally stable algorithms the second-phase price
is thus governed by $(N-\hat{N}_\pi)^2=O_p(N^2/n)$ under (A1), so that
\eqref{eq:degenerate-v2} is $O_p(N^2/n^2)$ --- asymptotically
negligible against a first-phase variance of order $N^2/n$. The same
closed form settles the bias question in this case without any
high-level condition: Remark~\ref{rem:degenerate-bias} in
Appendix~\ref{app:degenerate} shows that the degenerate estimator is
exactly $pq$-unbiased --- that is, unbiased under the joint $(p,q)$
randomization, the usage of \citet{sanguiao2021} --- under equal
first-phase probabilities, and has
relative bias $O(n^{-1})$ in general. The tree with no splits is the
case $H=1$ of the poststratified family of
Section~\ref{sec:properties-trees}, and the two routes agree: at
$H=1$ the second-phase share of Corollary~\ref{cor:relative-cost} is
of order $1/n$, so the partition is asymptotically free, which is
what part~(c) obtains directly from $V_q=O_p(N^2/n^2)$ against a
first-phase variance of order $N^2/n$. The substantive question,
answered in Section~\ref{sec:properties}, is how these conclusions
change when the algorithm's complexity grows.

\begin{remark}[Choice of the residual-correction sample]
\label{rem:variants}
Alternative constructions restrict the residual correction to a part
of $s$. The variant suggested by a literal transcription of the
classical two-phase estimator --- correction over $s_1$ only, with
two-phase weights $\pi^*_k=\pi_k\,n_1/n$ --- discards the observed
$y_k$ on $s_2$ and, in the degenerate case, inflates the variance by
a factor close to $1/f_1$; it is examined in the Supplementary
Material.

The construction of \citet{sanguiao2021} takes a third route, and
because its vocabulary overlaps with ours it is worth setting the
two side by side explicitly. Their single-split estimator (their
eq.~3.1) is
\begin{equation}
\label{eq:sanguiao}
\hat{Y}_1^{M}
=\sum_{k\in s_1}y_k
+\sum_{k\in U\setminus s_1}\hat{m}_1(\mathbf{x}_k)
+\sum_{k\in s_2}\frac{y_k-\hat{m}_1(\mathbf{x}_k)}{\pi_{2k}},
\qquad
\pi_{2k}=\Pr(k\in s_2\mid s_1),
\end{equation}
where the observed $y_k$ on $s_1$ enter uncorrected, prediction is
carried out on $U\setminus s_1$, and the residual correction runs
over $s_2$ only, weighted by the \emph{conditional} second-phase
inclusion probabilities $\pi_{2k}$ rather than by the first-phase
$\pi_k$. That choice of weights is what makes \eqref{eq:sanguiao}
\emph{exactly} $pq$-unbiased (their Prop.~6): the correction has zero
conditional expectation given $s_1$ by construction. They then
Rao--Blackwellize it, defining
$\hat{Y}^{*}_{M}=E_q(\hat{Y}_1^{M}\mid s)$ (their eq.~3.2), with a
Monte Carlo version averaging $K$ independent replicates of
\eqref{eq:sanguiao}.

The distinction we wish to draw is not one of quality but of
identity, since $\hat{Y}^{*}_{M}$ and the estimator $\hat{Y}_{RB}$ of
Section~\ref{sec:estimator-relations} below are both conditional
expectations over the partition and yet are different objects. They
are built from different base estimators --- \eqref{eq:sanguiao}
with weights $\pi_{2k}$ against \eqref{eq:ysl-collapsed} with weights
$\pi_k$ and the correction pooled over all of $s$ --- and the
Rao--Blackwell step plays a different role in each. In their
construction the base estimator is already exactly unbiased, so
averaging is pure variance reduction and nothing else; in ours the
base estimator is not exactly unbiased, and
Proposition~\ref{prop:rb-identity} shows that the average inherits
\emph{precisely the same} bias, which is therefore attributable to
the model-assisted form and not to the partition
(Section~\ref{sec:properties-bias}). The trade-off between the two
designs is correspondingly clean: \eqref{eq:sanguiao} buys exact
unbiasedness at the price of discarding the residual information in
$s_1$ and of weights that depend on the realized $s_1$, whereas
\eqref{eq:ysl-collapsed} uses every observed $y_k$ with the ordinary
design weights and pays with the standard model-assisted bias. These are two different base constructions with different
properties, not two attempts at the same one; what this paper studies
is the estimator produced by a single realized partition, which the
framework of \citet{sanguiao2021} does not analyze. For cell-mean
rules the two constructions are nevertheless linked by an exact
identity --- $\hat{Y}_{TS}$ is a convex combination of
\eqref{eq:sanguiao} and a poststratified plug-in
(Section~\ref{sec:composite}) --- and
Section~\ref{sec:discussion} returns to the systematic comparison.
\end{remark}

\subsection{An exact composite representation for cell-mean rules}
\label{sec:composite}

For cell-mean rules, the relation between the collapsed form
\eqref{eq:ysl-collapsed}, the single-split construction
\eqref{eq:sanguiao}, and the familiar full-fit estimator is not
merely structural: it is an exact identity. Let the algorithm
applied to $s_1$ partition the covariate space into cells
$U_1,\dots,U_H$ --- the partition may depend on $s_1$, as with a
tree grown on the training data --- and predict within each cell the
average of the fitted $y$-values,
$\hat{m}_1(\mathbf{x}_k)=\bar{y}_{s_{1h}}$ for $k\in U_h$. Write
\[
\hat{Y}_{P_1}
=\sum_{k\in U}\hat{m}_1(\mathbf{x}_k)
=\sum_{h=1}^{H}N_h\,\bar{y}_{s_{1h}}
\]
for the poststratified plug-in built on $s_1$.

\begin{proposition}[Composite representation]
\label{prop:composite}
Suppose the first phase has equal inclusion probabilities
$\pi_k\equiv n/N$ and that $q$ makes the conditional second-phase
inclusion probabilities constant,
$\Pr(k\in s_2\mid s_1)=\pi_2$ for all $k\in U\setminus s_1$ ---
as holds for $\mathrm{SRSWOR}$ at both phases, where
$\pi_2=n_2/(N-n_1)$ with $n_2=n-n_1$ by (P4), and for stratified
$\mathrm{SRSWOR}$ at both phases with proportional allocation and
$q$ stratified by the design strata. Then, for any cell-mean rule
fitted on $s_1$,
\begin{equation}
\label{eq:composite}
\hat{Y}_{TS}
=(1-c)\,\hat{Y}_{P_1}+c\,\hat{Y}_1^{M},
\qquad
c=\frac{N}{n}\,\pi_2=\frac{1-f_1}{1-f_1\,n/N}\in(0,1),
\end{equation}
with $\hat{Y}_1^{M}$ the single-split estimator \eqref{eq:sanguiao}.
Consequently, exactly and at every sample size,
\begin{equation}
\label{eq:composite-bias}
E(\hat{Y}_{TS})-Y=(1-c)\,\bigl\{E(\hat{Y}_{P_1})-Y\bigr\},
\end{equation}
since $\hat{Y}_1^{M}$ is exactly $pq$-unbiased.
\end{proposition}

The proof (Appendix~\ref{app:composite}) rests on a single fact:
cell means make the training residuals sum to zero within each cell,
$\sum_{k\in s_{1h}}\{y_k-\hat{m}_1(\mathbf{x}_k)\}=0$. Three
consequences organize its interpretation.

First, the pooled correction of \eqref{eq:ysl-collapsed} draws
nothing from $s_1$: writing
$e_k=y_k-\hat{m}_1(\mathbf{x}_k)$,
\[
\hat{Y}_{TS}-\hat{Y}_{P_1}=\frac{N}{n}\sum_{k\in s_2}e_k,
\qquad
\hat{Y}_1^{M}-\hat{Y}_{P_1}=\frac{N-n_1}{n_2}\sum_{k\in s_2}e_k ,
\]
so the two estimators correct the same plug-in with the same
out-of-sample residuals and differ \emph{only} in the weight given
to that correction: $\hat{Y}_{TS}$ undercorrects, relative to the
exactly unbiasing weight, by the factor $c$. In particular, the
leave-one-out variant of the point estimator --- replacing
$\hat{m}_1(\mathbf{x}_k)$, for $k\in s_1$, by its within-cell
leave-one-out version in the correction term --- changes nothing:
the leave-one-out factor $n_{1h}/(n_{1h}-1)$ multiplies within-cell
sums that vanish, so the point estimate is unchanged, although the
same device does change the \emph{variance} estimator
(Section~\ref{sec:varest-v1}).

Second, \eqref{eq:composite-bias} locates the design bias of the
single-partition estimator. Under $\mathrm{SRSWOR}$ at both phases,
$s_1$ is marginally an $\mathrm{SRSWOR}(n_1)$ sample from $U$
(fact~(P4)), so $\hat{Y}_{P_1}$ is distributed exactly as the
familiar poststratified (full-fit plug-in) estimator computed from a
probability sample of size $n_1$; the bias of $\hat{Y}_{TS}$ is the
fraction $1-c$ --- approximately $f_1$ when $n/N$ is small --- of
the bias of the estimator of common practice at the training-sample
size. The single-partition estimator interpolates between an exactly
unbiased but noisier construction ($c=1$ would give
$\hat{Y}_1^{M}$ itself) and the plug-in whose bias question,
by Proposition~\ref{prop:bias}, is the one already open for
partition-averaged estimators.

Third, the same vanishing-sum fact applied with $s_1=s$ shows that
the familiar full-fit estimator does not correct anything at all.

\begin{corollary}
\label{cor:fullfit-plugin}
For a cell-mean rule fitted on all of $s$, under
$\pi_k\equiv n/N$, the correction term of \eqref{eq:generic-ma}
vanishes identically:
$\sum_{k\in s}\{y_k-\hat{m}(\mathbf{x}_k)\}/\pi_k=0$, so the
full-fit model-assisted estimator reduces to the plug-in
$\sum_{k\in U}\hat{m}(\mathbf{x}_k)$.
\end{corollary}

The ``model-assisted'' form of the full fit is, for cell-mean rules
under equal-probability designs, cosmetic: the estimate rests
entirely on the fitted structure, and the protection that the
correction term is meant to provide has been fitted away. Training
on $s_1$ restores a correction with genuine content --- the weighted
out-of-sample residual sum above --- and
\eqref{eq:composite-bias} prices exactly how much of the plug-in
bias that correction removes. Section~\ref{sec:sim-point} verifies
both statements numerically.

Two remarks on scope. Under designs with unequal $\pi_k$ the
identity requires $\pi$-weighted cell means (under the
equal-probability designs above, weighted and unweighted means
coincide); with nonproportional allocation and $q$ stratified by the
design strata, \eqref{eq:composite} holds stratum by stratum, with
stratum-specific $c_g$, whenever the cells are nested within the
strata. And the identity concerns rules whose cells are built from
$s_1$. In the $s$-measurable configuration of
Remark~\ref{rem:adaptive}(ii) --- structure built on $s$, cell
means from $s_1$ --- the vanishing-sum fact still holds, but the
second component depends on $s$ beyond $s_1$ and is no longer
exactly unbiased, so \eqref{eq:composite-bias} does not transfer.

\subsection{Relation to partition averaging and cross-fitting}
\label{sec:estimator-relations}

The conditional expectation of $\hat{Y}_{TS}$ over the second phase
has a closed form. It identifies the object that appears inside the
variance decomposition of Section~\ref{sec:properties-decomp} ---
$V_1$ is by definition its design variance --- and it connects the
single-partition estimator to the partition-averaging family.

\begin{proposition}
\label{prop:rb-identity}
For any algorithm with $E_q\{|\hat{m}_1(\mathbf{x}_k)|\mid s\}<\infty$,
\begin{equation}
\label{eq:rb-estimator}
E_q(\hat{Y}_{TS}\mid s)
= \sum_{k\in U}\bar{m}_s(\mathbf{x}_k)
+ \sum_{k\in s}\frac{y_k-\bar{m}_s(\mathbf{x}_k)}{\pi_k}
\;=:\;\hat{Y}_{RB},
\end{equation}
the model-assisted estimator built on the partition-averaged rule
\eqref{eq:rb-rule}. Consequently: (a) $E(\hat{Y}_{TS})=E(\hat{Y}_{RB})$
--- the single-partition estimator and its partition-averaged version
have \emph{identical design bias}, exactly, for any algorithm, any
$(p,q)$, and any sample size; and (b) for the Monte Carlo average
$\hat{Y}^{(B)}=B^{-1}\sum_{b=1}^{B}\hat{Y}_{TS}^{(b)}$ over $B$
independent draws $s_1^{(b)}\sim q(\cdot\mid s)$,
$E_q(\hat{Y}^{(B)}\mid s)=\hat{Y}_{RB}$ for every $B$.
\end{proposition}

Proposition~\ref{prop:rb-identity}(a) sharpens the discussion of
Section~\ref{sec:intro}: averaging over partitions is \emph{pure
variance reduction}. The averaging cannot repair bias, and
conversely the single partition costs no bias --- its entire price is
the second-phase variance quantified in
Section~\ref{sec:properties}. The family
$\{\hat{Y}^{(B)}:B\geq 1\}$ interpolates between the estimator studied
here ($B=1$) and the idealized partition-averaged estimator
($B\to\infty$), at a computational cost of $B$ refits; the Monte Carlo
subsampling Rao--Blackwell estimator of \citet{sanguiao2021} is the
$B$-replicate average of \emph{their} single-split term, arranged for
exact $pq$-unbiasedness as described in Remark~\ref{rem:variants},
with the same averaging device. Theorem~\ref{thm:decomp} below turns
this interpolation into a single exact variance formula.

The rule $\bar{m}_s$ is not computable in general --- it is an
expectation over all partitions --- and it never has to be. Here
$\hat{Y}_{RB}$ is identified, not estimated, and the variance
estimators of Section~\ref{sec:varest} are computed from the
realized fit alone.

The relation to cross-fitting is structural rather than algebraic. The
$K$-fold cross-fitted estimator
\citep{dagdoug2026,kwon2026} partitions $U$ at random into $K$
folds, and hence $s$, and matches
each unit's residual with a prediction fitted on the folds excluding
it; when the folds are of equal size in $s$, each fold pair resembles
a single-partition construction with training fraction $(K-1)/K$ and
correction restricted to the held-out fold --- with folds formed in
$U$ before sampling, the realized fractions are random --- and the
sum over folds reassembles a full-sample correction with out-of-fold
residuals. Three differences separate that
object from \eqref{eq:ysl-collapsed}: the cross-fitted estimator
combines $K$ complementary partitions rather than realizing one; its
training fraction is tied to $K$ rather than free; and the fold assignment is held fixed, with its effect treated as
asymptotically negligible, whereas here the randomness of the
partition is the object of study. We return to the
comparison in Section~\ref{sec:properties-trees}, where the
second-phase variance identifies the regime in which fold randomness
ceases to be negligible.

\section{Design-based properties}
\label{sec:properties}

\subsection{Design bias}
\label{sec:properties-bias}

By Proposition~\ref{prop:rb-identity}, the bias of $\hat{Y}_{TS}$ is
the bias of $\hat{Y}_{RB}$, a model-assisted estimator with the
$s$-measurable rule $\bar{m}_s$. Its bias therefore has the standard
model-assisted structure.

\begin{proposition}
\label{prop:bias}
(a) Exactly, for any algorithm and design,
\begin{equation}
\label{eq:bias-exact}
E(\hat{Y}_{TS})-Y
= -\sum_{k\in U}\mathrm{Cov}_p\!\left(\frac{I_k}{\pi_k},\,
\bar{m}_s(\mathbf{x}_k)\right).
\end{equation}
(b) Under (A1) and (A4),
\[
\frac{1}{N}\,\bigl|E(\hat{Y}_{TS})-Y\bigr|
\;\leq\;
\Bigl(\frac{N}{\lambda n}\Bigr)^{1/2}\delta_n .
\]
\end{proposition}

Part (a) shows that bias arises solely because the partition-averaged
rule co-varies with sample membership --- the same mechanism as for
any model-assisted estimator with an estimated rule
\citep{breidt2017} --- and part (b) gives the standard high-level
sufficient condition for its asymptotic negligibility. The bound is
a product of two factors: $(N/\lambda n)^{1/2}$ is the worst-case
standard deviation of the design weight $I_k/\pi_k$ permitted by
(A1), and $\delta_n$ measures how far the partition-averaged rule
still is from its deterministic limit --- bias per unit at most how
wild the weights may be times how unstable the algorithm is. The
first factor is unbounded: writing $\kappa_n=n/N$ it equals
$(\lambda \kappa_n)^{-1/2}$, so the bound vanishes only when
$\delta_n=o(\kappa_n^{1/2})$, a condition that tightens as the
sampling fraction shrinks. Under $n/N\to\kappa>0$ it is enough that
$\delta_n\to 0$; under $n/N\to 0$, the usual survey regime, more is
required of the algorithm. What part~(b) delivers is therefore
negligibility of the bias \emph{per population unit}; negligibility
\emph{relative to the standard error}, which is what inference
needs, would require $\delta_n=o(N^{-1/2})$. Sharper
statements (bias negligible relative to the standard error) follow
from (A4$'$) by the usual first-order-equivalence argument, and
establishing (A4$'$) for adaptive learners is the open frontier
discussed in Section~\ref{sec:intro}; we do not claim progress on it
here. What \emph{is} new is the reduction itself: by
Proposition~\ref{prop:rb-identity}, the bias question for the
single-partition estimator is \emph{identical} to the bias question
for partition-averaged estimators. Whatever bias theory holds for the
averaged estimator transfers verbatim, and the analysis of the
partition can legitimately concentrate on variance, as we now do.

\subsection{Exact variance under the joint randomization}
\label{sec:properties-decomp}

\begin{theorem}
\label{thm:decomp}
Let $V_1=V_p(\hat{Y}_{RB})$ and $V_2=V_q(T_1\mid s)
=V_q\{\sum_{k\in U}(1-I_k/\pi_k)\,\hat{m}_1(\mathbf{x}_k)\mid s\}$.
Then, exactly, for any algorithm and any $(p,q)$:
\begin{align}
V(\hat{Y}_{TS}) &= V_1 + E_p(V_2),
\label{eq:main-decomp}\\
V(\hat{Y}^{(B)}) &= V_1 + \frac{1}{B}\,E_p(V_2),
\qquad B=1,2,\ldots,
\label{eq:family}
\end{align}
where $\hat{Y}^{(B)}$ averages $B$ independent second-phase draws as
in Proposition~\ref{prop:rb-identity}(b), and \eqref{eq:main-decomp}
is the case $B=1$.
\end{theorem}

Three comments. First, \eqref{eq:main-decomp} is an exact identity,
not an approximation: the derivation of Section~\ref{sec:estimator}
places all second-phase randomness in $T_1$, so the
decomposition \eqref{eq:tp-decomp} applies with $V_2$ depending only
on the training fit. It is in this sense that the variance is
\emph{derived} for the present estimator rather than inherited from
the classical formulas, which do not cover it
(Section~\ref{sec:setup-classical}).

Second, \eqref{eq:family} embeds the single-partition estimator and
the partition-averaging family in one formula. The idealized
partition-averaged estimator is the limit $B\to\infty$, with variance
$V_1$; the Monte Carlo version with $B$ refits retains $E_p(V_2)/B$;
the estimator actually reported after one fit has the full
$E_p(V_2)$.

\begin{corollary}
\label{cor:cost}
The design price of a single partition, relative to idealized
partition averaging, is exactly
$V(\hat{Y}_{TS})-V_p(\hat{Y}_{RB})=E_p(V_2)$, and relative to a
$B$-replicate average it is $(1-1/B)\,E_p(V_2)$. In relative terms,
the proportion of the variance of $\hat{Y}_{TS}$ attributable to the
second phase is $E_p(V_2)/\{V_1+E_p(V_2)\}$.
\end{corollary}

Third, Corollary~\ref{cor:cost} makes precise the question posed in
Section~\ref{sec:intro} --- how much precision does a single partition
cost --- and identifies $E_p(V_2)$ as the quantity to be evaluated. It
also identifies what a fixed-prediction analysis misses: an analysis
that treats the fitted predictions as fixed addresses (at best) the
first-phase component, and understates the variance of the reported
estimator by the second-phase share. The remainder of this section
evaluates $V_2$: exactly in the degenerate case (already done in
Proposition~\ref{prop:degenerate}(c)), and exactly up to explicit
remainders for tree-type predictors
(Section~\ref{sec:properties-trees}). The structure
$V_1+E_p(V_2)/B$ is formally the within/between decomposition of
multiple imputation \citep{rubin1987}, with the partition draw in
the role of the imputation draw; the analogy is exact at the level
of Theorem~\ref{thm:decomp}, although the inferential targets
differ.

\subsection{First-phase approximate variance}
\label{sec:properties-v1}

The first-phase component $V_1=V_p(\hat{Y}_{RB})$ is the variance of a
model-assisted estimator with an $s$-measurable rule, and its
approximation is classical.

\begin{proposition}
\label{prop:av1}
Under (A1), (A2), (A4$'$),
\begin{equation}
\label{eq:av1}
V_1 = \sum_{k\in U}\sum_{\ell\in U}
\Delta_{k\ell}\,\frac{E_k}{\pi_k}\,\frac{E_\ell}{\pi_\ell}
\;+\; o\!\left(\frac{N^2}{n}\right),
\qquad E_k=y_k-m_\nu(\mathbf{x}_k).
\end{equation}
\end{proposition}

We emphasize the epistemic status of \eqref{eq:av1}: it is the usual
model-assisted approximate variance, valid under the usual high-level
conditions, and it is \emph{not} where this paper claims novelty. For
the tree case below we do not rely on it; instead $V_1$ is evaluated
directly through the poststratification structure, which is what makes
the comparison of the two phases sharp.

\subsection{Second-phase variance for tree-type predictors}
\label{sec:properties-trees}

We now specialize to prediction rules with poststratification
structure. Let $\{A_1,\ldots,A_H\}$ be a partition of the covariate
space into $H$ cells, and let the training fit predict the training
mean of the cell:
\begin{equation}
\label{eq:ps-rule}
\hat{m}_1(\mathbf{x})=\bar{y}_{s_{1h}}
\quad\text{for }\mathbf{x}\in A_h,
\end{equation}
with the convention $\hat{m}_1(\mathbf{x})=\bar{y}_{s_1}$ on cells
with empty training intersection. Fitted regression trees have
exactly this form, with cells given by the terminal nodes
\citep{mcconville2019}. Throughout this subsection the partition is
assumed \emph{$s$-measurable}: fitted from the full sample $s$, from
frame information, or fixed in advance. The fully adaptive case, in
which the cell structure itself is refitted on each $s_1$, is
discussed in Remark~\ref{rem:adaptive}.

Define, for each cell, the population and sample intersections and
their sizes,
\[
U_h=U\cap A_h,\qquad
s_h=s\cap A_h,\qquad
s_{1h}=s_1\cap A_h,\qquad
s_{2h}=s_2\cap A_h,
\]
\[
N_h=\#U_h,\qquad
n_h=\#s_h,\qquad
n_{1h}=\#s_{1h},\qquad
\hat{N}_{\pi,h}=\sum_{k\in s_h}\frac{1}{\pi_k},\qquad
D_h=N_h-\hat{N}_{\pi,h},
\]
where $D_h$ is the cell-level HT count error, computable because
$N_h$ is known from the frame, and $S^2_{y,s,h}$ is the sample
variance of $y$ over $s_h$. The cells $A_h$ are constructions of the
prediction rule, not the design strata of the first phase. Under
\eqref{eq:ps-rule} the second-phase term collapses to a weighted sum
of training cell means,
$\sum_{k\in U}(1-I_k/\pi_k)\hat{m}_1(\mathbf{x}_k)
=\sum_{h}D_h\,\bar{y}_{s_{1h}}$, and its conditional variance can
be evaluated.

\begin{theorem}
\label{thm:trees}
Let the partition be $s$-measurable and the training fit be
\eqref{eq:ps-rule}.

(i) If the second phase is stratified by cell --- in the sense of
the prediction rule's terminal nodes, not of the first-phase design
strata --- drawing
$n_{1h}\geq 1$ units by SRSWOR within each $s_h$ independently
across cells, then exactly
\begin{equation}
\label{eq:v2-trees-exact}
V_2 = \sum_{h=1}^{H} D_h^2
\left(\frac{1}{n_{1h}}-\frac{1}{n_h}\right) S^2_{y,s,h}.
\end{equation}

(ii) If the second phase is global $\mathrm{SRSWOR}(n_1)$ from $s$
(condition (A3)), then, conditionally on the event
$\Omega_0=\{n_{1h}\geq 1\ \forall h\}$,
\begin{equation}
\label{eq:v2-trees-global}
V_q(T_1\mid s,\Omega_0)
= \sum_{h=1}^{H} D_h^2
\left\{E_q\!\left(\frac{1}{n_{1h}}\,\Big|\;\Omega_0\right)
-\frac{1}{n_h}\right\} S^2_{y,s,h},
\end{equation}
exactly, where $n_{1h}$ is hypergeometric with mean $n_1n_h/n$;
moreover $\Pr(\Omega_0^c\mid s)\leq H(1-f_1)^{n_{\min}}$ with
$n_{\min}=\min_h n_h$, and if $f_1 n_{\min}\geq 2$,
\[
V_2 = \frac{1-f_1}{f_1}\sum_{h=1}^{H}
D_h^2\,\frac{S^2_{y,s,h}}{n_h}\,\{1+r_h\},
\qquad |r_h|\leq \frac{C_2}{f_1 n_h},
\]
for an absolute constant $C_2$.
\end{theorem}

Part~(i) covers the case in which the cells are fixed in advance,
determined by frame information, or fitted on $s$, and the training
subsample is drawn stratified by cell; part~(ii) covers global
$\mathrm{SRSWOR}$, the configuration used in the adaptive
simulations of Section~\ref{sec:simulation}. In part~(ii) the
conditioning on $\Omega_0$ concerns the exact hypergeometric form,
not the definition of $V_2$: under the convention below
\eqref{eq:ps-rule} --- a cell with empty training intersection
predicts $\bar{y}_{s_1}$ --- the unconditional
$V_2=V_q(T_1\mid s)$ is well defined, it is the quantity
Theorem~\ref{thm:decomp} uses, and the bound
$\Pr(\Omega_0^c\mid s)\leq H(1-f_1)^{n_{\min}}$ makes the
difference between the two immaterial at the cell sizes used here.

Formula \eqref{eq:v2-trees-exact} has a transparent reading: each cell
contributes the subsampling variance of its training mean, scaled by
the squared cell-level count error of the first phase,
$D_h=N_h-\hat{N}_{\pi,h}$; the factor
$(1-f_1)/f_1$ carries the entire dependence on the training fraction, and
$H=1$ recovers Proposition~\ref{prop:degenerate}(c). All ingredients
of \eqref{eq:v2-trees-exact} are computable from the observed sample
and the frame, a point exploited in Section~\ref{sec:varest}.

The payoff is the order of the second phase relative to the first as
the number of cells grows.

\begin{corollary}
\label{cor:relative-cost}
Let the first phase be $\mathrm{SRSWOR}(n)$ with $n/N\to\kappa<1$,
let the second phase be stratified by cell as in
Theorem~\ref{thm:trees}(i) with $n_{1h}=\lceil f_1n_h\rceil$,
let the partition be fixed with $H=H_n$ cells satisfying
$c_3\,n/H\leq n_h\leq c_4\,n/H$ and
$c_3\,N/H\leq N_h\leq c_4\,N/H$ for constants $0<c_3\leq c_4$, let
$fn/H\to\infty$, and let the within-cell variances satisfy
$c_5\leq S^2_{y,U,h}/\bar{S}^2\leq c_6$ for constants
$0<c_5\leq c_6$, where $\bar{S}^2=H^{-1}\sum_h S^2_{y,U,h}>0$. Then
there exist constants $0<c_\ast\leq C_\ast<\infty$, depending only
on $c_3,\dots,c_6$ and $\kappa$, such that
\begin{equation}
\label{eq:relative-cost}
c_\ast\,\frac{1-f_1}{f_1}\cdot\frac{H}{n}
\;\leq\;
\frac{E_p(V_2)}{V_1}
\;\leq\;
C_\ast\,\frac{1-f_1}{f_1}\cdot\frac{H}{n}.
\end{equation}
\end{corollary}

Under the global second phase of (A3) the same order holds up to
the remainders $r_h$ of Theorem~\ref{thm:trees}(ii) and the
exponentially rare event $\Omega_0^c$; the concentration steps use
the tail bounds of \citet[Sec.~6]{hoeffding1963} and
\citet{serfling1974} (Appendix~\ref{app:relative-cost}).

Corollary~\ref{cor:relative-cost} is, in our view, the central
qualitative finding of the paper, for it separates two regimes.
If $H$ is fixed --- a stable, low-complexity predictor --- the
second-phase share is $O(1/n)$ and the partition is asymptotically
free, consistent with the degenerate case and with the classical
comfort of poststratification. If instead $H=H_n\asymp n^{\alpha}$
grows with the sample, with $\alpha<1$ so that the regime condition
$f_1n/H\to\infty$ holds, the share is of order $n^{\alpha-1}$: it
vanishes, but arbitrarily slowly as $\alpha$ approaches one, and at
any given $n$ it grows with the number of cells. The boundary case
$\alpha=1$ --- terminal nodes of bounded size, the regime of deeply
grown trees --- lies \emph{outside} the hypotheses of the corollary,
and the direction of the failure is known: when cells are small the
rounding $n_{1h}=\lceil f_1n_h\rceil$ pushes the within-cell
training fraction toward one, so \eqref{eq:relative-cost}
overstates the share there. We therefore record it as a conjecture,
not a result, that under bounded node size the second-phase share
remains bounded away from zero; a proof would need a different
argument for $1/n_{1h}-1/n_h$ and for the empty-cell event at
bounded $n_h$. A fixed-prediction analysis understates the variance
of the reported estimator by the second-phase share in every regime;
within the corollary's regime that understatement is governed by
$\{(1-f_1)/f_1\}(H/n)$. This identifies one design-based
mechanism, distinct from the shrinkage of in-sample residuals and
additive to it, behind the variance underestimation documented
empirically for flexible learners by \citet[Sec.~6.2]{dagdoug2023}
and \citet{cosenza2025}, and characterized theoretically, for the
high-dimensional linear GREG, by \citet{bouhadra2026};
Section~\ref{sec:sim-variance} suggests that in the adaptive regime
it is the smaller of the two. It also locates the premise of the
cross-fitting variance estimators: treating the effect of the
partition as asymptotically negligible rests on the out-of-fold
predictions being mean-square consistent for a population-level
rule (\citealp[eq.~(15)]{dagdoug2026};
\citealp[assumption~(A5)]{kwon2026}), a condition compatible with
complexity that grows with the sample size --- including linear
models of growing dimension, as in \citet{kwon2026} --- but not
with complexity proportional to it. Whether the fold-assignment
share of the cross-fitted estimator's variance is negligible at
practical sample sizes is an empirical question, made measurable by
Theorem~\ref{thm:decomp} --- the fold assignment is a $q$ --- and
left to a separate study.
The corollary also quantifies the role of the training fraction: the
second-phase share scales as $(1-f_1)/f_1$, decreasing in $f_1$, making
$f_1$ --- the central experimental variable of the simulation study in
Section~\ref{sec:simulation} --- an explicit design parameter whose variance price is known in
closed form when the cell structure is fixed or $s$-measurable.
The rounding mechanism just described is the same integrality issue
flagged in \eqref{eq:rounding}, applied cell by cell, and it is the
source of the remainders $r_h$ of Theorem~\ref{thm:trees}(ii); it
is also a caveat for simulation design, where cell sizes must be
kept large enough for the asymptotic reading to apply.

\begin{remark}[Adaptive cell structure]
\label{rem:adaptive}
When the tree structure is itself refitted, the results above depend
on it only through which sample determined it, and three cases
separate.

(i) \emph{Structure fixed in advance or determined by the frame.}
The cells are not random, and Theorem~\ref{thm:trees},
Corollary~\ref{cor:relative-cost} and
Proposition~\ref{prop:v2-known} apply as stated; the analytic
estimator is exact and requires no refits.

(ii) \emph{Structure fitted on the full sample $s$.} The cells are
$s$-measurable: random before $s$ is drawn, constant once $s$ is
known. That is all the conditioning argument requires, so the
conclusions of (i) hold verbatim. This case is implementable rather
than merely hypothetical --- grow the structure on $s$ and compute
the cell means on $s_1$ alone --- and it combines a
\citet{mcconville2019} structure with out-of-sample cell means; it
is the survey counterpart of ``honest'' trees
\citep{athey2016,wager2018}, with the roles of the two halves
reversed. Section~\ref{sec:simulation} runs it.

(iii) \emph{Structure refitted on each training subsample.} The
cells are $s_1$-measurable and the conditioning argument behind
Theorem~\ref{thm:trees} no longer applies. This is the case of a
tree grown directly on $s_1$; $s_1$-measurability is a statement
about which data determined the cells, not about whether the
algorithm randomizes internally. It is the configuration run in
Section~\ref{sec:simulation}.

What survives in (iii) is more than what is lost.
Theorem~\ref{thm:decomp} is algorithm-free and continues to budget
the full second-phase variance, structure randomness included.
Proposition~\ref{prop:rep-unbiased} holds for any algorithm and any
$A\geq 2$, so $\hat{V}_2^{\mathrm{rep}}$ remains exactly
conditionally unbiased: since each replicate refits the structure,
the dispersion of the $T_1^{*(a)}$ absorbs that randomness without
modelling it. What (iii) lacks is the closed form available when
the cell structure is fixed or $s$-measurable, and the order of
Corollary~\ref{cor:relative-cost} --- a formula, not an estimator.

The simulations of Section~\ref{sec:simulation} measure what that
costs. With trees refitted on each training subsample the empirical
second-phase share exceeds the fixed-structure benchmark by an
order of magnitude, which locates the dominant component of $V_2$
in the randomness of the structure itself rather than in the
variability of the cell means --- a quantity that, to our
knowledge, has not previously been measured. It also fixes the
status of the conditional expression: it is a lower bound on the
second-phase cost, and therefore still informative in (iii), as a
floor.

The randomness of the structure is a second route by which the same
partition enters, not a separate problem. Whether the leading order
of \eqref{eq:relative-cost} survives when the structure is refitted
under a minimum-node-size condition of the type used by
\citet{mcconville2019}, with the structure operating on the same
cells-of-size-$n/H$ scale, is a question the simulations do not
settle: they use \texttt{rpart} defaults, whose minimum node size
does not grow with $n$, and in that configuration the order of
magnitude above the benchmark suggests that structure randomness
enters at a different scale. We therefore leave the order of the
adaptive case open rather than conjecture it; the results above
should be read as conditional on the realized structure, the
reading legitimized by the poststratification interpretation of
\citet{mcconville2019}.
\end{remark}

\begin{remark}[Random forests]
\label{rem:forests}
A forest averages trees grown on resamples of the training set, adding
algorithmic randomization to the second phase; design-based variance
theory for bagged estimators under complex designs is a long-standing
open problem \citep{wang2014}. Theorem~\ref{thm:decomp} still applies
--- \eqref{eq:main-decomp} is algorithm-free, with the algorithmic
randomization absorbed into $V_2$ (or averaged out if the forest is
viewed as approximating its own expectation) --- but we make no
analytic claim for forests beyond it. Forests are evaluated empirically in
Section~\ref{sec:simulation}. The replication estimator
\eqref{eq:v2rep} transfers to them unchanged, since
Proposition~\ref{prop:rep-unbiased} is algorithm-free; the first-phase
component does not, and the leave-one-out residuals \eqref{eq:loo} are
replaced there by their out-of-bag analogue.
\end{remark}

\section{Variance estimation}
\label{sec:varest}

Guided by Theorem~\ref{thm:decomp}, we estimate the two components
separately, $\hat{V}=\hat{V}_1+\hat{V}_2$, and we require of any
proposal that it leave the reported point estimate --- the realized
single-partition $\hat{Y}_{TS}$ --- untouched.

\subsection{First-phase component}
\label{sec:varest-v1}

The first-phase component is estimated by the standard
Horvitz--Thompson-type residual formula with unit weight
($g_k\equiv 1$),
\begin{equation}
\label{eq:v1hat}
\hat{V}_1
= \sum_{k\in s}\sum_{\ell\in s}
\frac{\Delta_{k\ell}}{\pi_{k\ell}}\,
\frac{\hat{e}_k}{\pi_k}\,\frac{\hat{e}_\ell}{\pi_\ell},
\qquad
\hat{e}_k = y_k-\hat{m}_1(\mathbf{x}_k),
\end{equation}
with the Sen--Yates--Grundy form available for fixed-size designs. The
residuals in \eqref{eq:v1hat} raise, in miniature, the issue that
motivates the paper: for $k\in s_1$ they are in-sample residuals of
the training fit and may be shrunken by adaptive fitting, while for
$k\in s_2$ they are genuinely out-of-sample. The two groups pull in
opposite directions relative to the target: $V_1$ is the variance of
the estimator built on the averaged rule $\bar{m}_s$, while
\eqref{eq:v1hat} uses residuals from the noisier realized fit
$\hat{m}_1$ --- larger than residuals from $\bar{m}_s$ would be on
$s_2$, in-sample and smaller on $s_1$ --- so the sign of the net
first-phase bias is an empirical matter, reported directly against
$V_1$ in Section~\ref{sec:sim-variance}. For tree-type predictors
a cheap correction is available: replace, for $k\in s_{1h}$, the
residual by its leave-one-out version within the cell,
\begin{equation}
\label{eq:loo}
\hat{e}^{\,\mathrm{loo}}_k
= \frac{n_{1h}}{n_{1h}-1}\,(y_k-\bar{y}_{s_{1h}}),
\end{equation}
which requires no refitting because leave-one-out means of cell
averages are closed-form. The factor
$n_{1h}/(n_{1h}-1)$ is the leverage correction for a cell-mean fit;
for the linear GREG in high-dimensional regimes,
\citet{bouhadra2026} show that the analogous leave-one-out
correction removes the asymptotic bias of the plug-in variance
estimator. Both versions of \eqref{eq:v1hat} are
assessed in Section~\ref{sec:simulation}; no claim of superiority is
made in advance.

\subsection{Analytic second-phase component for tree-type predictors}
\label{sec:varest-v2-analytic}

For poststratified rules, Theorem~\ref{thm:trees} makes the
second-phase component computable rather than merely estimable.

\begin{proposition}
\label{prop:v2-known}
Let the partition be $s$-measurable and the training fit be
\eqref{eq:ps-rule}, and define
\begin{equation}
\label{eq:v2hat}
\hat{V}_2
= \sum_{h=1}^{H} D_h^2
\left(\frac{1}{n_{1h}}-\frac{1}{n_h}\right) S^2_{y,s,h},
\qquad D_h=N_h-\hat{N}_{\pi,h}.
\end{equation}
Every ingredient of \eqref{eq:v2hat} is computable from $(s,s_1)$ and
the frame. Under the cell-stratified second phase of
Theorem~\ref{thm:trees}(i), $\hat{V}_2$ is $s$-measurable and
$\hat{V}_2=V_2$ exactly: the second-phase variance is \emph{known},
not estimated. Under global $\mathrm{SRSWOR}(n_1)$, conditionally on
$\Omega_0$, $E_q(\hat{V}_2\mid s,\Omega_0)=V_q(T_1\mid s,\Omega_0)$
exactly.
\end{proposition}

The exactness statement is a consequence of the structural reversal of
Section~\ref{sec:setup-classical}: because the second-phase design is
chosen and its ingredients ($N_h$ from the frame, $\hat{N}_{\pi,h}$,
$S^2_{y,s,h}$ from the full first-phase sample) are observed, nothing
about the second phase needs to be estimated. In the classical
two-phase setting the analogous quantity is not available, since $y$
is unobserved outside the second-phase sample.

\subsection{Replication estimator for general algorithms}
\label{sec:varest-v2-rep}

When the predictor lacks poststratification structure (forests,
boosting), $V_2$ is no longer available in closed form, but it remains
the conditional variance of a computable statistic under a
\emph{known} randomization mechanism, and can therefore be estimated
by direct replication of the second phase: draw
$s_1^{*(1)},\ldots,s_1^{*(A)}$ independently from $q(\cdot\mid s)$,
refit the algorithm on each, compute
$T_1^{*(a)}=\sum_{k\in U}(1-I_k/\pi_k)\,
\hat{m}_{s_1^{*(a)}}(\mathbf{x}_k)$, and set
\begin{equation}
\label{eq:v2rep}
\hat{V}_2^{\mathrm{rep}}
= \frac{1}{A-1}\sum_{a=1}^{A}
\bigl(T_1^{*(a)}-\bar{T}_1^{*}\bigr)^2,
\qquad
\bar{T}_1^{*}=\frac{1}{A}\sum_{a=1}^{A}T_1^{*(a)}.
\end{equation}

\begin{proposition}
\label{prop:rep-unbiased}
For any algorithm and any $A\geq 2$,
$E_q\{\hat{V}_2^{\mathrm{rep}}\mid s\}=V_2$ exactly, and
$E\{\hat{V}_2^{\mathrm{rep}}\}=E_p(V_2)$, the second-phase component
of \eqref{eq:main-decomp}.
\end{proposition}

Two design features matter here. First, \eqref{eq:v2rep} does not
alter the inferential object: the reported estimate remains the
realized-partition $\hat{Y}_{TS}$, and the replicate draws serve only
to measure the dispersion that the realized draw was subject to. This
is the operational difference from Monte Carlo Rao--Blackwellization,
where the replicate fits \emph{define} the point estimator
\citep{sanguiao2021}; a parallel replication device is used for
variance purposes in their framework as well. Second, the Monte Carlo
error of \eqref{eq:v2rep} decays as $1/A$ and $A$ multiplies only the
variance estimation cost, not the estimation cost; moderate values
(tens of replicates) suffice in the simulations of
Section~\ref{sec:simulation}.

One further point deserves emphasis, because the division of labour
between the two components of the variance estimator is easy to
misread. Equation~\eqref{eq:v2rep} targets $V_2$ alone: it measures
the dispersion induced by the second-phase randomization with $s$ held
fixed, and it neither addresses nor claims to address the dependence
between a unit and its own prediction, which is a first-phase matter
handled by the leave-one-out residuals \eqref{eq:loo}. The distinction
matters because replication devices that resample the \emph{data}
rather than the second phase carry no analogous guarantee:
\citet{ferreira2026}, in a weighting-pipeline setting with a learner
fitted on the same sample whose residuals the estimator then corrects,
reports that re-fitting the learner inside a bootstrap replicate
recovers only about two thirds of the true variance --- precisely
because the unit remains in the training set of nearly every replicate
--- and that cross-fitting is what repairs it. Under \eqref{eq:v2rep}
the replicate draws come from the known design $q(\cdot\mid s)$ and
the unit is outside the training subsample with probability $1-f_1$ by
construction, so Proposition~\ref{prop:rep-unbiased} applies as
stated; but the two devices answer different questions, and a
first-phase correction suited to the learner remains necessary in
either case.

\subsection{Combined estimator and intervals}
\label{sec:varest-combined}

The proposed variance estimator is
$\hat{V}=\hat{V}_1+\hat{V}_2$ with $\hat{V}_2$ given by
\eqref{eq:v2hat} for tree-type predictors and by \eqref{eq:v2rep} in
general. Its second-phase component is exactly conditionally unbiased
(Propositions~\ref{prop:v2-known} and~\ref{prop:rep-unbiased}); its
first-phase component carries the usual properties, and the usual
caveats, of residual-based model-assisted variance estimators
\citep{breidt2017}, including the in-sample shrinkage issue that
\eqref{eq:loo} addresses. Interval estimation in
Section~\ref{sec:simulation} uses the working normal approximation
$\hat{Y}_{TS}\pm z_{1-\alpha/2}\hat{V}^{1/2}$; we prove no central
limit theorem here (see Section~\ref{sec:discussion}), and the
adequacy of the approximation is assessed empirically, in line with
the coverage assessments of \citet{dagdoug2023} and
\citet{kwon2026}.

The computational budgets can be stated in terms of algorithm
fits, typically the dominant cost with flexible learners. Point
estimation requires: one fit for $\hat{Y}_{TS}$; $B$ fits for the
$B$-replicate partition average (with $B$ in the tens to hundreds
in practice; \citealp{sanguiao2021,zhang2026}); $K$ fits for
$K$-fold cross-fitting, whose out-of-fold variance estimator then
needs no further fits \citep{dagdoug2026,kwon2026}. Variance
estimation for $\hat{Y}_{TS}$ adds no fit when the analytic
tree-case $\hat{V}_1+\hat{V}_2$ of
\eqref{eq:v1hat}--\eqref{eq:v2hat} applies --- cell structure
fixed or $s$-measurable, a single fit in all --- and adds $A$
refits for \eqref{eq:v2rep} when the structure is adaptively
refitted, for a total of $1+A$ fits, against $K$ for cross-fitting
and $B$-plus-further-work for the partition average, whose refits
deliver the point estimator but not its variance. Which budget buys
which coverage at matched fit counts is the comparison deferred to
a separate study; Corollary~\ref{cor:relative-cost} prices what
the single partition itself costs in precision, a trade-off
quantified over the training fraction $f_1$ in
Section~\ref{sec:simulation}.

\section{Simulation study}
\label{sec:simulation}

\subsection{Design}
\label{sec:sim-design}

We generated three finite populations of size $N=10{,}000$, modeled on
the populations of \citet{dagdoug2023}. Auxiliary variables were drawn
once and held fixed: $x_0\sim U(0,1)$, and standardized
$x_2\sim\mathrm{Beta}(3,1)$, $x_3\sim\Gamma(3,2)$,
$x_6\sim\mathrm{Exp}(1)$. The three study variables are
\emph{linear}, $y=1+2(x_0-0.5)+\varepsilon$ with
$\varepsilon\sim N(0,0.1^2)$; \emph{nonlinear},
$y=2+(x_6+x_2+x_3)^2+\varepsilon$ with $\varepsilon\sim N(0,0.1^2)$,
which combines skewness, curvature and interactions; and \emph{weak
signal}, the linear mean function with $\varepsilon\sim N(0,1)$
(population $R^2\approx 0.25$), the scenario in which flexible fitting
has little to offer, so that whatever the partition costs is not
recovered.

Three configurations are run. In the \emph{adaptive}
configuration --- the main one --- the first phase is simple random
sampling without replacement with $n=500$ and the second phase is
global $\mathrm{SRSWOR}(n_1)$ from $s$ with training fractions
$f_1\in\{0.5,0.7,0.9\}$, the central experimental variable. Regression trees are fitted with
\texttt{rpart} at its default settings ($\mathit{cp}=0.01$, minimum
split size 20), so the tree structure is refitted on each realized
$s_1$ --- the fully adaptive case of Remark~\ref{rem:adaptive}, which
the theory covers only conditionally; the simulation is precisely the
check that the unconditional behavior matches. The returned tree is
a deterministic function of $s_1$: the structure is fixed by the
growing rule at $\mathit{cp}=0.01$, and no pruning step based on
\texttt{rpart}'s internal cross-validation is applied. The
second-phase randomness is therefore exactly that of $q$. Random
forests differ in this respect --- \texttt{ranger} resamples $s_1$
and draws its candidate variables at random --- and that additional
randomization is absorbed into $V_2$, as discussed in
Remark~\ref{rem:forests}; the forests themselves use
\texttt{ranger} with 100 trees and default \texttt{mtry}, and
each replicate fit inside \eqref{eq:v2rep} draws a fresh forest
seed, so that $\hat{V}_2^{\mathrm{rep}}$ includes the seed
variability that the reported fit also carries
(Appendix~\ref{app:rep-unbiased}).

In the \emph{$s$-measurable} configuration
(Remark~\ref{rem:adaptive}(ii)) the tree structure is grown on all
of $s$ with the same \texttt{rpart} defaults, the cell means are
computed on $s_1$ alone, and $q$ is stratified by the realized
cells with $n_{1h}=\lceil f_1n_h\rceil$ --- the configuration in
which Theorem~\ref{thm:trees}(i), Corollary~\ref{cor:relative-cost}
and Proposition~\ref{prop:v2-known} are exact. It is run at
$(n,N)=(500,10^4)$ and $(2{,}000,4\times 10^4)$, the larger
population generated by the same mechanism, so that the sampling
fraction stays at $n/N=0.05$ and the order claim of
Corollary~\ref{cor:relative-cost} is tested across $n$. In the
\emph{stratified} configuration the first phase is stratified
$\mathrm{SRSWOR}$ over five equal strata defined by quintiles of
$x_0$ (linear and weak-signal populations) or $x_6$ (nonlinear),
with proportional allocation $n_g=100$, and $q$ is stratified by
the same strata with $n_{1g}=f_1n_g$ --- the setting of
Proposition~\ref{prop:composite}, whose identity the scripts verify
at every replicate (largest deviation across all runs:
$4\times 10^{-11}$, machine precision at this scale). Under
$\mathrm{SRSWOR}$ the degenerate case has zero second-phase
variance (Proposition~\ref{prop:degenerate}), so everything the
adaptive simulations measure enters through the cell-count errors
$D_h$; the stratified configuration moves those errors, which is
its point. We use
$R=1{,}000$ Monte Carlo replicates per population (Monte Carlo
standard error for a 95\% coverage estimate: about $0.7$ percentage
points; for an RRMSE, at most $0.1$ percentage points, and
considerably less for differences between estimators, which are
computed on the same samples), with fixed seeds; all results in this section are produced by
the scripts in the supplementary material and every number in the
tables is written by those scripts, not transcribed.

The estimators compared are: the Horvitz--Thompson estimator; the
linear GREG on the covariates of the population's mean function; the
full-fit tree-assisted estimator $\hat{Y}_{\mathrm{tree}}$ (fitted on
all of $s$, in-sample correction), which under $\mathrm{SRSWOR}$ is
the tree-based poststratification estimator $\sum_h N_h\bar{y}_{s_h}$
of \citet{mcconville2019}, with the tree grown by \texttt{rpart} as
above rather than by their algorithm; the same full-fit estimator
with the tree grown under the node-size rule of their theory,
$k(n)=n^{11/20}$ (minimum node size $31$ and minimum split size
$62$ at $n=500$, other settings unchanged --- their significance
test is omitted, as by their own account it does not enter the
proofs), labelled ``node rule'' in the tables; the proposed
single-partition estimator $\hat{Y}_{TS}^{\mathrm{tree}}$ at each
$f_1$; the exactly unbiased single-split estimator
$\hat{Y}_1^{M}$ of \eqref{eq:sanguiao} at each $f_1$; the
$B$-replicate partition average $\hat{Y}^{(B)}$ of
Proposition~\ref{prop:rb-identity}(b) with $B=50$ at $f_1=0.7$,
included as a measuring device rather than as a competitor
(Section~\ref{sec:sim-share}); the
$K$-fold cross-fitted tree estimator ($K=5$), with out-of-fold
residuals and fold-averaged census predictions; and the random-forest
versions $\hat{Y}_{\mathrm{rf}}$ (full fit) and
$\hat{Y}_{TS}^{\mathrm{rf}}$ ($f_1=0.7$), reported as an
illustration only. For variance estimation we
evaluate, for $\hat{Y}_{TS}^{\mathrm{tree}}$ at each $f_1$, the analytic
estimator $\hat{V}_1^{\mathrm{loo}}+\hat{V}_2$ of
Section~\ref{sec:varest} (leave-one-out residuals \eqref{eq:loo}
within $s_1$, cells taken from the realized tree); at $f_1=0.7$ also the
replication version $\hat{V}_1^{\mathrm{loo}}+\hat{V}_2^{\mathrm{rep}}$
($A=30$) and three benchmarks: $\hat{V}_1^{\mathrm{loo}}$ alone,
which accounts for the sampling of $s$ but not for the partition;
the same first-phase term computed from raw in-sample residuals,
$\hat{V}_1$ alone; and the na\"ive variance of
$\hat{Y}_{\mathrm{tree}}$ with in-sample residuals,
\[
\hat{V}^{\mathrm{naive}}
= N^2\Bigl(\frac{1}{n}-\frac{1}{N}\Bigr)\frac{1}{n-1}
\sum_{k\in s}(e_k-\bar{e})^2,
\qquad e_k=y_k-\hat{m}_s(\mathbf{x}_k),
\]
with $\hat{m}_s$ the tree fitted on all of $s$ --- the estimator
whose downward bias motivates the paper. It is the same sandwich as
$\hat{V}_1$; what differs is that the residuals are in-sample
rather than corrected, and that no second-phase term is added.
Under $\mathrm{SRSWOR}$ it is the plug-in variance estimator of
\citet[Sec.~2]{mcconville2019}, whose consistency (their
Theorem~3) requires every terminal node to hold at least $k(n)$
sample units, with $k(n)$ growing faster than $n^{1/2}$
($n^{11/20}\approx 31$ at $n=500$ in their implementation). The
\texttt{rpart} defaults used here fix the minimum terminal-node
size at $7$ units whatever the sample size, so the results for this
estimator concern a configuration outside those conditions and do
not bear on their theorem; the node-rule rows provide the
complementary configuration, inside their conditions. For the
forest we evaluate
$\hat{V}_1^{\mathrm{oob}}+\hat{V}_2^{\mathrm{rep}}$ ($A=15$; residuals
within $s_1$ replaced by out-of-bag residuals) in the nonlinear
population. For the cross-fitted estimator we also evaluate its
own out-of-fold variance estimator --- the same sandwich as
$\hat{V}_1$, with out-of-fold residuals in place of in-sample ones
\citep{dagdoug2026,kwon2026} --- since cross-fitting is the
alternative most often proposed for the same task; variance
estimation for the partition-averaged estimator is not evaluated,
as it accompanies a different reported number. For the full-fit
estimator under the node rule, the na\"ive variance is evaluated
with its own residuals.
Performance measures are percent relative bias (RB) and
percent relative root mean squared error (RRMSE) for point estimators;
for variance estimators, percent relative bias against the Monte Carlo
variance of the corresponding point estimator, empirical coverage of
the nominal 95\% normal interval, and mean interval length as a
percentage of $Y$.

Two deliberate reductions relative to a full factorial deserve
notice. First, unequal-probability ($\pi$ps) first-phase designs
are not included; the scripts accept them, and nothing in the
estimators is specific to equal probabilities, but under unequal
probabilities the composite identity requires $\pi$-weighted cell
means and the closed forms would need rederivation, an evaluation
left to a separate study. Second, the $B$-replicate average is
computed for the tree only: at $B=50$ refits per replicate it is
by far the most expensive point estimator in the comparison, which
is part of the point.

\subsection{Point estimation}
\label{sec:sim-point}

\begin{table}[t]
\centering
\caption{Point estimators: percent relative bias, percent relative
root mean squared error, and mean squared error relative to the GREG
($=100$), by population. $R=1{,}000$ replicates, $n=500$. The
partition average $\hat{Y}^{(B)}$, below the rule, is the reference
quantity used in Section~\ref{sec:sim-share} to isolate $E_p(V_2)$,
not a competitor. The ``node rule'' row grows the tree under the
minimum-node-size rule $k(n)=n^{11/20}$ of \citet{mcconville2019};
$\hat{Y}_1^{M}$ is the exactly unbiased single-split estimator
\eqref{eq:sanguiao}. Monte Carlo standard errors of RRMSE entries
are at most $0.1$ points; paired standard errors for differences
quoted in the text are smaller.}
\label{tab:point}
\small
\begin{tabular}{l rrr rrr rrr}
\toprule
& \multicolumn{3}{c}{Linear} & \multicolumn{3}{c}{Nonlinear}
& \multicolumn{3}{c}{Weak signal} \\
\cmidrule(lr){2-4}\cmidrule(lr){5-7}\cmidrule(lr){8-10}
Estimator & RB & RRMSE & MSE & RB & RRMSE & MSE & RB & RRMSE & MSE \\
\midrule
HT & -0.05 & 2.59 & 3413 & 0.05 & 4.40 & 128 & 0.09 & 4.98 & 127 \\
GREG & 0.00 & 0.44 & 100 & -0.28 & 3.89 & 100 & 0.14 & 4.42 & 100 \\
$\hat{Y}_{\mathrm{tree}}$ & -0.01 & 0.62 & 195 & -1.24 & 3.51 & 81 & 0.17 & 4.50 & 104 \\
$\hat{Y}_{\mathrm{tree}}$ (node rule) & -0.01 & 0.62 & 195 & -0.35 & 3.75 & 93 & 0.16 & 4.48 & 103 \\
$\hat{Y}_{TS}^{\mathrm{tree}}$ ($f_1=0.5$) & -0.01 & 0.64 & 205 & -0.51 & 3.62 & 86 & 0.12 & 4.66 & 111 \\
$\hat{Y}_{TS}^{\mathrm{tree}}$ ($f_1=0.7$) & -0.02 & 0.65 & 212 & -0.88 & 3.48 & 80 & 0.09 & 4.57 & 107 \\
$\hat{Y}_{TS}^{\mathrm{tree}}$ ($f_1=0.9$) & -0.03 & 0.63 & 200 & -1.07 & 3.47 & 80 & 0.15 & 4.54 & 106 \\
$\hat{Y}_{1}^{M}$ ($f_1=0.5$) & 0.02 & 0.87 & 383 & 0.07 & 4.85 & 155 & -0.06 & 6.43 & 212 \\
$\hat{Y}_{1}^{M}$ ($f_1=0.7$) & -0.02 & 1.18 & 701 & -0.03 & 6.07 & 243 & -0.11 & 8.44 & 365 \\
$\hat{Y}_{1}^{M}$ ($f_1=0.9$) & -0.07 & 1.91 & 1860 & 0.42 & 10.40 & 713 & 0.31 & 13.80 & 977 \\
CF tree ($K=5$) & -0.01 & 0.63 & 199 & 0.08 & 3.56 & 84 & 0.14 & 4.53 & 105 \\
$\hat{Y}_{\mathrm{rf}}$ & 0.01 & 0.50 & 125 & -1.29 & 2.58 & 44 & 0.07 & 4.92 & 124 \\
$\hat{Y}_{TS}^{\mathrm{rf}}$ ($f_1=0.7$) & -0.00 & 0.51 & 130 & -1.02 & 2.68 & 48 & 0.05 & 4.98 & 127 \\
\midrule
\multicolumn{10}{l}{\emph{Reference quantity (Section~\ref{sec:sim-share})}}\\
$\hat{Y}^{(B)}$ ($B=50$, $f_1=0.7$) & -0.01 & 0.55 & 152 & -0.84 & 3.19 & 67 & 0.16 & 4.48 & 103 \\
 
\bottomrule
\end{tabular}
\end{table}

\begin{figure}[t]
\centering
\includegraphics[width=\textwidth]{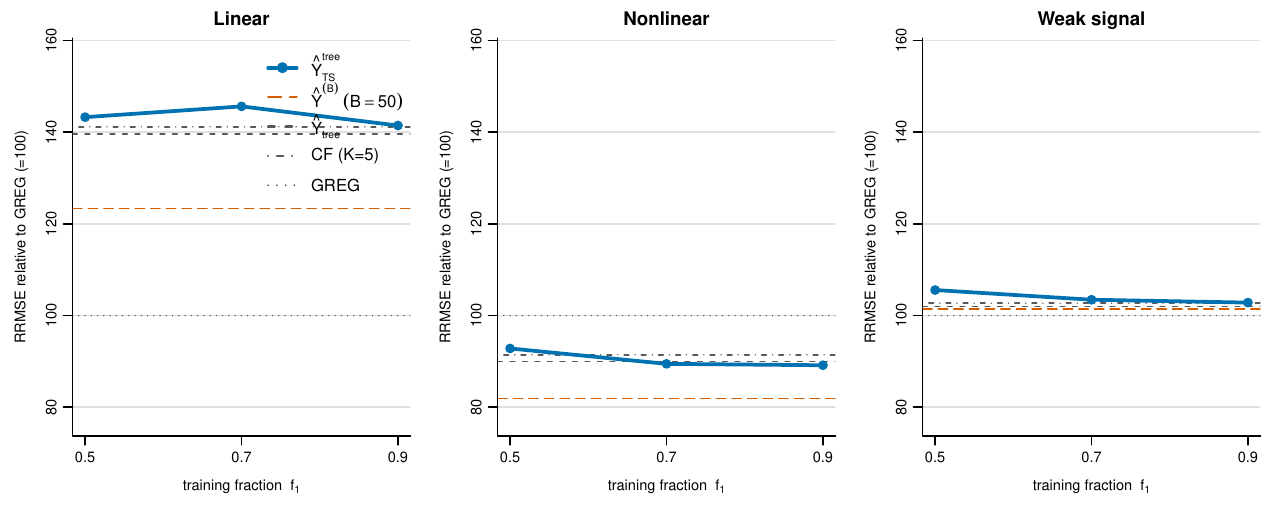}
\caption{RRMSE of the single-partition tree estimator as a function of
the training fraction $f_1$ (solid line), against its partition-averaged
version $\hat{Y}^{(B)}$, the full-fit tree estimator, the $K$-fold
cross-fitted estimator, and the GREG (horizontal references).
$R=1{,}000$ replicates.}
\label{fig:efficiency}
\end{figure}

Table~\ref{tab:point} and Figure~\ref{fig:efficiency} give the point
results. In the linear and weak-signal populations all point
estimators are essentially design-unbiased, with relative biases
below $0.2\%$. In the nonlinear population every estimator that
fits a flexible learner without cross-fitting carries a material
bias, discussed below. The three populations tell
three complementary stories. In the \emph{linear}
population the GREG dominates everything (RRMSE $0.44\%$ against
$0.62$--$0.65\%$ for the tree-based estimators): flexible fitting buys
nothing when the working model is right, and the single partition
costs at most $0.03$ percentage points of RRMSE relative to the
full-fit tree --- a relative penalty under $5\%$ --- while the
$B=50$ partition average recovers part of it ($0.55\%$). In the
\emph{nonlinear} population the ordering reverses: every tree- and
forest-assisted estimator beats the GREG, and the single-partition
estimator at $f_1=0.7$ and $f_1=0.9$ (RRMSE $3.48\%$ and $3.47\%$)
matches the full-fit tree ($3.51\%$): the paired differences,
$-0.02$ and $-0.03$ percentage points, are within their Monte Carlo
standard errors ($0.06$ and $0.04$).

The bias column is where Proposition~\ref{prop:composite} can be
watched operating. The full fit shows $-1.24\%$ --- and by
Corollary~\ref{cor:fullfit-plugin} this is pure plug-in bias, since
its correction term is identically zero. The single-split rows are
unbiased at every $f_1$, exactly as
Proposition~\ref{prop:composite} requires, and the single-partition
rows sit at the fraction $1-c$ of the plug-in's bias at the
training-sample size: with $n/N=0.05$, $1-c$ is $0.49$, $0.69$ and
$0.90$ at $f_1=0.5$, $0.7$ and $0.9$, and applying
\eqref{eq:composite-bias} to the observed bias of
$\hat{Y}_{P_1}$ predicts $-0.55\%$, $-0.87\%$ and $-1.12\%$,
against $-0.51\%$, $-0.88\%$ and $-1.07\%$ observed --- agreement
within Monte Carlo error, with no asymptotics involved. The
single-split rows also price exact unbiasedness: their RRMSE rises
from $4.85\%$ at $f_1=0.5$ to $10.40\%$ at $f_1=0.9$, where the
assessment sample is smallest, against $3.47$--$3.62\%$ for
$\hat{Y}_{TS}$; the three constructions differ only in the weight
their correction gives the same out-of-sample residuals --- $0$ for
the plug-in, $c$ for $\hat{Y}_{TS}$, $1$ for the single split ---
and the bias--variance exchange along that line is what the table
displays. The forest shows the same pattern ($-1.02\%$ with the
partition at $f_1=0.7$, $-1.29\%$ with the full fit). The node-rule
row adds the other lever: grown under $k(n)=n^{11/20}$, the
full-fit tree's bias falls to $-0.35\%$ at a small RRMSE cost
($3.75\%$ against $3.51\%$) --- consistent with the role their
theory assigns to a growing minimum node size.

Among the flexible-learner estimators, cross-fitting attains the
smallest bias relative to its standard error ($+0.08\%$, about
$0.02$ standard errors) at comparable RRMSE ($3.56\%$), at the
cost of $K=5$ fits; its own variance estimator is evaluated in
Section~\ref{sec:sim-variance} alongside ours.

The random forest attains the best RRMSE ($2.58\%$), at the price
of the largest bias, a point that resurfaces in the coverage
results. In the
\emph{weak-signal} population, where the auxiliary information leaves little for a
flexible fit to exploit, assisted estimators cluster
around the GREG ($4.42\%$) or above it, the single-partition
estimator at $f_1=0.5$ pays a visible price ($4.66\%$), and nothing
recovers the cost of flexible fitting.

Table~\ref{tab:stratified} repeats the point comparison under the
stratified first phase. The orderings are unchanged: the stratified
design tightens the linear population (RRMSE of
$\hat{Y}_{TS}^{\mathrm{tree}}$ at $f_1=0.7$: $0.59\%$ against
$0.65\%$ under $\mathrm{SRSWOR}$) and moves the others little; the
single-split estimator remains exactly unbiased --- its largest
entry, $+0.64\%$ in the nonlinear population, is Monte Carlo noise:
a $20{,}000$-replicate check of that cell returns $-0.03\%$ with a
standard error of $0.04$ --- and the composite identity holds at
machine precision in every replicate, as the design section
records. The bias of $\hat{Y}_{TS}$ again tracks
$(1-c)$ times the plug-in bias, per stratum-preserving $q$.

\begin{table}[t]
\centering
\caption{Stratified first phase (proportional allocation, $q$
stratified by the design strata): percent relative bias and RRMSE.
$R=1{,}000$ replicates, $n=500$, $n_g=100$ per stratum. The
composite identity $\hat{Y}_{TS}=(1-c)\hat{Y}_{P_1}
+c\hat{Y}_1^{M}$ is verified at machine precision in every
replicate.}
\label{tab:stratified}
\small
\begin{tabular}{l rr rr rr}
\toprule
& \multicolumn{2}{c}{Linear} & \multicolumn{2}{c}{Nonlinear}
& \multicolumn{2}{c}{Weak signal} \\
\cmidrule(lr){2-3}\cmidrule(lr){4-5}\cmidrule(lr){6-7}
Estimator & RB & RRMSE & RB & RRMSE & RB & RRMSE \\
\midrule
HT & -0.02 & 0.65 & 0.14 & 4.38 & -0.05 & 4.55 \\
GREG & 0.01 & 0.43 & -0.05 & 3.85 & -0.03 & 4.54 \\
$\hat{Y}_{\mathrm{tree}}$ & -0.01 & 0.58 & -0.96 & 3.45 & -0.01 & 4.60 \\
$\hat{Y}_{TS}^{\mathrm{tree}}$ ($f_1=0.5$) & 0.01 & 0.58 & -0.46 & 3.59 & 0.02 & 4.68 \\
$\hat{Y}_{TS}^{\mathrm{tree}}$ ($f_1=0.7$) & -0.02 & 0.59 & -0.66 & 3.56 & 0.00 & 4.63 \\
$\hat{Y}_{TS}^{\mathrm{tree}}$ ($f_1=0.9$) & -0.01 & 0.58 & -0.92 & 3.51 & -0.01 & 4.60 \\
$\hat{Y}_{1}^{M}$ ($f_1=0.7$) & -0.04 & 1.09 & 0.64 & 5.99 & 0.31 & 8.21 \\
CF tree ($K=5$) & -0.01 & 0.61 & 0.28 & 3.59 & -0.00 & 4.68 \\
 
\bottomrule
\end{tabular}
\end{table}

In every population $\hat{Y}^{(B)}$ has a smaller mean squared
error than $\hat{Y}_{TS}^{\mathrm{tree}}$ at $f_1=0.7$ ($152$
against $212$, $67$ against $80$ and $103$ against $107$), and it
must: its bias is exactly that of the single-partition estimator
(Proposition~\ref{prop:rb-identity}(a)) and its variance is smaller
by $(1-1/B)E_p(V_2)$ (Theorem~\ref{thm:decomp}), so no run could
order them otherwise. For the same reason averaging cannot improve
the ratio of bias to standard error: it leaves the bias unchanged
and shrinks the standard error ($0.26$ and $0.27$ standard errors
in the nonlinear population). An analyst who can afford $B$ refits
and does not need the reported number to come from one fitted rule
should average. The setting of this paper is the other one: an
estimate reported from a single realized fit --- for correspondence
between estimate and inference, for computation, or for
transparency, the three reasons of Section~\ref{sec:intro} ---
whose variance statement must then include the second phase.

\subsection{The $s$-measurable configuration: where the theory is
exact}
\label{sec:sim-smedible}

\begin{table}[t]
\centering
\caption{$s$-measurable configuration (structure grown on $s$, cell
means on $s_1$, $q$ stratified by cell): RRMSE of
$\hat{Y}_{TS}^{\mathrm{tree}}$, empirical second-phase share
$E_p(V_2)/V(\hat{Y}_{TS})$ with the analytic $V_2$ (exact here),
its benchmark $\{(1-f_1)/f_1\}(\bar{H}/n)$, and relative bias and
coverage of $\hat{V}_1^{\mathrm{loo}}+\hat{V}_2$. $R=1{,}000$
replicates; $n/N=0.05$ at both sample sizes.}
\label{tab:smeasurable}
\small
\begin{tabular}{ll r rrr rr}
\toprule
Population & $n$ & $f_1$ & RRMSE & Share & Benchmark
& RB$(\hat{V})$ & Cov. \\
& & & (\%) & (\%) & (\%) & (\%) & (\%) \\
\midrule
Linear & 500 & 0.5 & 0.62 & 0.82 & 1.29 & -11.7 & 93.8 \\
 &  & 0.7 & 0.62 & 0.34 & 0.55 & -13.2 & 93.0 \\
 &  & 0.9 & 0.62 & 0.09 & 0.14 & -13.5 & 93.4 \\
\addlinespace
Linear & 2000 & 0.5 & 0.29 & 0.28 & 0.35 & -2.2 & 96.1 \\
 &  & 0.7 & 0.29 & 0.12 & 0.15 & -2.0 & 95.5 \\
 &  & 0.9 & 0.29 & 0.03 & 0.04 & -1.7 & 95.7 \\
\addlinespace
Nonlinear & 500 & 0.5 & 3.49 & 4.33 & 2.26 & -11.7 & 86.8 \\
 &  & 0.7 & 3.46 & 1.89 & 0.97 & -15.8 & 85.6 \\
 &  & 0.9 & 3.44 & 0.21 & 0.25 & -19.0 & 84.8 \\
\addlinespace
Nonlinear & 2000 & 0.5 & 1.77 & 2.90 & 0.81 & -20.3 & 81.5 \\
 &  & 0.7 & 1.76 & 1.26 & 0.35 & -20.8 & 81.7 \\
 &  & 0.9 & 1.75 & 0.17 & 0.09 & -22.9 & 80.8 \\
\addlinespace
Weak signal & 500 & 0.5 & 4.62 & 0.62 & 0.92 & -10.0 & 93.7 \\
 &  & 0.7 & 4.60 & 0.26 & 0.39 & -9.9 & 93.6 \\
 &  & 0.9 & 4.60 & 0.06 & 0.10 & -10.2 & 93.6 \\
\addlinespace
Weak signal & 2000 & 0.5 & 2.23 & 0.14 & 0.20 & -2.5 & 94.7 \\
 &  & 0.7 & 2.23 & 0.06 & 0.09 & -2.5 & 94.4 \\
 &  & 0.9 & 2.23 & 0.02 & 0.02 & -2.6 & 94.4 \\
 
\bottomrule
\end{tabular}
\end{table}

The $s$-measurable configuration is where
Theorem~\ref{thm:trees}(i) and Proposition~\ref{prop:v2-known} are
exact, and Table~\ref{tab:smeasurable} runs it. Three facts stand
out. First, the variance theory behaves exactly as stated: the
analytic $\hat{V}_2$ and the replication estimator agree to the
reported precision in every cell --- they estimate the same
quantity, one in closed form at no refitting cost --- and the
second-phase share follows the benchmark
$\{(1-f_1)/f_1\}(\bar{H}/n)$ of
Corollary~\ref{cor:relative-cost} in level and in order: within a
factor of two everywhere (the constants $c_\ast,C_\ast$ absorb the
cell-size heterogeneity), decreasing in $f_1$ as $(1-f_1)/f_1$, and
falling with $n$ at fixed $n/N$ --- at $f_1=0.7$ the share moves
from $0.34$, $1.89$ and $0.26\%$ at $n=500$ to $0.12$, $1.26$ and
$0.06\%$ at $n=2{,}000$, tracking the decline of $\bar{H}/n$.
Second, where the point estimator is essentially unbiased ---
linear and weak-signal populations --- the intervals behave: at
$n=2{,}000$ coverage is $95.5$--$96.1\%$ and $94.4$--$94.7\%$
with variance biases within $\pm 3\%$; at $n=500$ coverage sits
at $93.0$--$93.8\%$, the shortfall coming from the finite-sample
downward bias of the first-phase term ($-10$ to $-14\%$), not from
the second phase. Third, and it is the honest half of the result:
in the nonlinear population this configuration inherits the
\emph{full-fit structure bias} --- the cells are grown on all of
$s$, so the point estimator is biased by $-1.26\%$, essentially
the plug-in bias of Table~\ref{tab:point} --- and coverage falls
to $81$--$87\%$ for a reason that has nothing to do with variance
estimation. The two configurations thus split the difficulty
between them: the $s$-measurable one has exact, free variance
estimation and a tiny second-phase share, but keeps the full-fit
bias; the adaptive one reduces that bias by the factor $1-c$ of
Proposition~\ref{prop:composite}, and pays with the structure
randomness measured next.

\subsection{The empirical second-phase share}
\label{sec:sim-share}

Theorem~\ref{thm:decomp} makes the second-phase component observable
without relying on any variance estimator. By \eqref{eq:family},
$V(\hat{Y}_{TS})-V(\hat{Y}^{(B)})=(1-1/B)\,E_p(V_2)$ exactly, so the
Monte Carlo variances of the two point estimators isolate $E_p(V_2)$
without using any of the variance estimators evaluated in
Section~\ref{sec:sim-variance}. This is the role of $\hat{Y}^{(B)}$
in the study: it is not a competitor but the empirically available
approximation to $\hat{Y}_{RB}$, whose variance is the first-phase
component. With $B=50$ the factor $1-1/B=0.98$ is immaterial at the
precision reported. Comparing $V(\hat{Y}_{TS})$ at $f_1=0.7$ with
$V(\hat{Y}^{(50)})$, the empirical share
$E_p(V_2)/\{V_1+E_p(V_2)\}$ is about $17\%$ in the nonlinear
population --- the only one in which the tree-based estimators
improve on the GREG (Table~\ref{tab:point}) --- and ranges from
$4\%$ in the weak-signal population to $28\%$ in the linear one,
with Monte Carlo standard errors of $2.2$, $1.2$ and $2.6$ points.
A second measurement uses the replication estimator itself: by
Proposition~\ref{prop:rep-unbiased} its average over the replicates
estimates $E_p(V_2)$ without bias, and it gives shares of $23\%$,
$18.5\%$ and $4\%$ in the linear, nonlinear and weak-signal
populations, with standard errors of $1.1$, $0.9$ and $0.2$ points.
The two measurements agree within Monte Carlo error, which is also
an empirical check of Proposition~\ref{prop:rep-unbiased}, and
together they place the share at about $17$--$19\%$ in the
nonlinear population and between $23$ and $29\%$ in the linear one.
Both are an order of magnitude larger than in the $s$-measurable
configuration, where the same instruments measured shares of
$0.3$, $1.9$ and $0.3\%$ at $f_1=0.7$
(Table~\ref{tab:smeasurable}), and than the fixed-structure
benchmark $\{(1-f_1)/f_1\}(H/n)$ evaluated at the observed average
tree sizes ($\bar{H}=6.4$, $9.7$ and $5.7$, giving
$0.5$--$0.8\%$). The gap between the two configurations is a
direct measurement of the randomness of the \emph{tree structure
itself}: with \texttt{rpart} refitted on each $s_1$, structure
randomness contributes roughly twenty times the share that cell-mean
randomness contributes with the structure held $s$-measurable. The
exact decomposition still budgets it --- it is algorithm-free ---
but this gap between the adaptive and the conditional regimes,
anticipated in Remark~\ref{rem:adaptive}, has direct consequences
for variance estimation.

\subsection{Variance estimation and coverage}
\label{sec:sim-variance}

\begin{table}[tp]
\centering
\caption{Variance estimators for the single-partition estimator:
percent relative bias against the Monte Carlo variance, empirical
coverage of the nominal 95\% interval, and mean interval length as a
percentage of $Y$. $R=1{,}000$ replicates, $n=500$.}
\label{tab:variance}
\small
\begin{tabular}{l rrr}
\toprule
Variance estimator & RB (\%) & Coverage (\%) & Length (\%$Y$) \\
\midrule
\multicolumn{4}{l}{\emph{Linear population}}\\
$\hat{V}_1^{\mathrm{loo}}+\hat{V}_2$ ($f_1=0.5$) & -10.4 & 93.7 & 2.36 \\
$\hat{V}_1^{\mathrm{loo}}+\hat{V}_2$ ($f_1=0.7$) & -16.5 & 91.8 & 2.31 \\
$\hat{V}_1^{\mathrm{loo}}+\hat{V}_2$ ($f_1=0.9$) & -14.3 & 93.4 & 2.27 \\
$\hat{V}_1^{\mathrm{loo}}+\hat{V}_2^{\mathrm{rep}}$ ($f_1=0.7$, $A=30$) & 6.0 & 95.3 & 2.60 \\
$\hat{V}_1^{\mathrm{loo}}$ only ($f_1=0.7$) & -16.9 & 91.7 & 2.31 \\
$\hat{V}_1$ with raw residuals only ($f_1=0.7$) & -18.7 & 91.6 & 2.28 \\
CF tree ($K=5$), out-of-fold $\hat{V}$ & -0.6 & 95.0 & 2.45 \\
$\hat{Y}_{\mathrm{tree}}$, na\"ive $\hat{V}$ & -15.8 & 92.3 & 2.22 \\
$\hat{Y}_{\mathrm{tree}}$ (node rule), na\"ive $\hat{V}$ & -15.8 & 92.3 & 2.22 \\
\addlinespace
\multicolumn{4}{l}{\emph{Nonlinear population}}\\
$\hat{V}_1^{\mathrm{loo}}+\hat{V}_2$ ($f_1=0.5$) & -5.3 & 91.2 & 13.48 \\
$\hat{V}_1^{\mathrm{loo}}+\hat{V}_2$ ($f_1=0.7$) & -8.4 & 89.5 & 12.47 \\
$\hat{V}_1^{\mathrm{loo}}+\hat{V}_2$ ($f_1=0.9$) & -17.5 & 86.7 & 11.59 \\
$\hat{V}_1^{\mathrm{loo}}+\hat{V}_2^{\mathrm{rep}}$ ($f_1=0.7$, $A=30$) & 8.1 & 92.2 & 13.57 \\
$\hat{V}_1^{\mathrm{loo}}$ only ($f_1=0.7$) & -10.4 & 89.0 & 12.34 \\
$\hat{V}_1$ with raw residuals only ($f_1=0.7$) & -20.4 & 87.7 & 11.64 \\
CF tree ($K=5$), out-of-fold $\hat{V}$ & -8.7 & 92.8 & 13.18 \\
$\hat{Y}_{\mathrm{tree}}$, na\"ive $\hat{V}$ & -33.7 & 82.4 & 10.31 \\
$\hat{Y}_{\mathrm{tree}}$ (node rule), na\"ive $\hat{V}$ & -10.4 & 92.0 & 13.72 \\
RF: $\hat{V}_1^{\mathrm{oob}}+\hat{V}_2^{\mathrm{rep}}$ ($f_1=0.7$, $A=15$) & 14.9 & 90.2 & 10.29 \\
\addlinespace
\multicolumn{4}{l}{\emph{Weak signal population}}\\
$\hat{V}_1^{\mathrm{loo}}+\hat{V}_2$ ($f_1=0.5$) & -7.9 & 93.4 & 17.53 \\
$\hat{V}_1^{\mathrm{loo}}+\hat{V}_2$ ($f_1=0.7$) & -7.6 & 93.8 & 17.21 \\
$\hat{V}_1^{\mathrm{loo}}+\hat{V}_2$ ($f_1=0.9$) & -7.6 & 93.1 & 17.10 \\
$\hat{V}_1^{\mathrm{loo}}+\hat{V}_2^{\mathrm{rep}}$ ($f_1=0.7$, $A=30$) & -4.2 & 94.4 & 17.53 \\
$\hat{V}_1^{\mathrm{loo}}$ only ($f_1=0.7$) & -8.0 & 93.8 & 17.18 \\
$\hat{V}_1$ with raw residuals only ($f_1=0.7$) & -10.0 & 93.7 & 16.99 \\
CF tree ($K=5$), out-of-fold $\hat{V}$ & -0.1 & 93.8 & 17.75 \\
$\hat{Y}_{\mathrm{tree}}$, na\"ive $\hat{V}$ & -8.2 & 92.9 & 16.91 \\
$\hat{Y}_{\mathrm{tree}}$ (node rule), na\"ive $\hat{V}$ & -6.3 & 93.7 & 16.98 \\
 
\bottomrule
\end{tabular}
\end{table}

\begin{figure}[t]
\centering
\includegraphics[width=\textwidth]{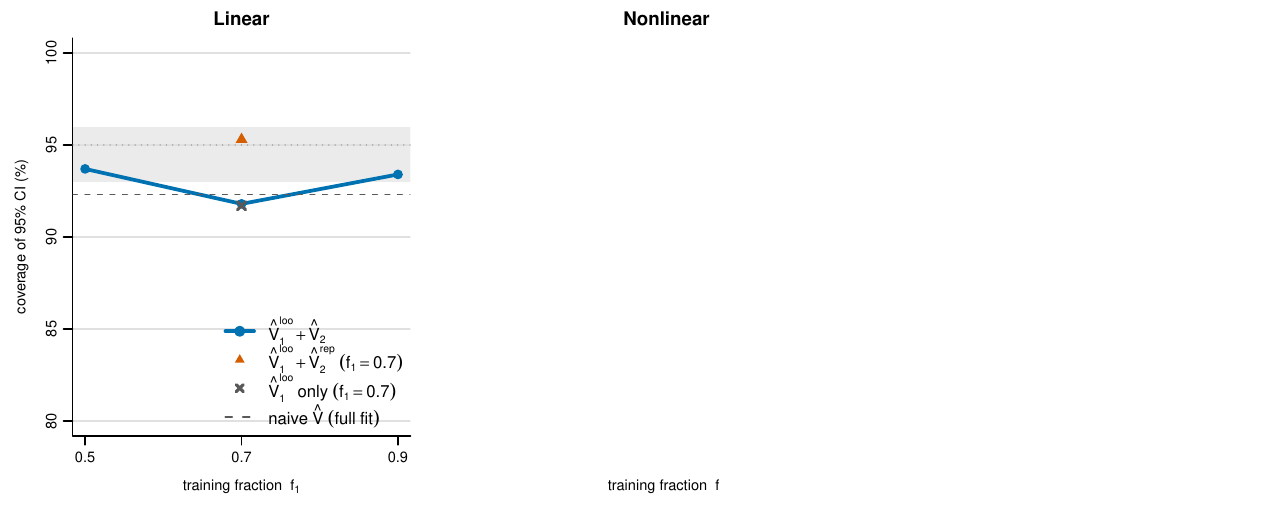}
\caption{Empirical coverage of nominal 95\% intervals for the
single-partition tree estimator with the proposed variance estimator
(solid line), against $\hat{V}_1^{\mathrm{loo}}$ alone and the na\"ive
variance of the full-fit tree estimator at $f_1=0.7$. The shaded band
marks 93--96\%. $R=1{,}000$ replicates.}
\label{fig:coverage}
\end{figure}

Table~\ref{tab:variance} and Figure~\ref{fig:coverage} evaluate
variance estimation against two baselines that answer different
questions. The first is the full-fit pipeline,
$\hat{Y}_{\mathrm{tree}}$ with its na\"ive variance. It covers
$92.3\%$ and $92.9\%$ in the linear and weak-signal populations,
where a tree is not the appropriate tool, and fails where it is: in
the nonlinear population, the one in which the tree genuinely
adapts, the na\"ive variance underestimates by $33.7\%$ and the
nominal 95\% interval covers $82.4\%$ of the time, against
$92.2\%$ for the single partition with the replication estimator.
That gap of $9.8$ points compares pipelines, and mixes a change of
point estimator with a change of variance estimator. The second
baseline isolates the second phase: $\hat{V}_1^{\mathrm{loo}}$
alone, computed for the same point estimator and with the
leave-one-out correction, accounts for the sampling of $s$ but not
for the partition. Along that same point estimator the ladder can
be attributed step by step: in the nonlinear population coverage
goes from $87.7\%$ with raw in-sample residuals to $89.0\%$ with
the leave-one-out correction and $92.2\%$ with the second phase, so
that of the $9.8$ points separating the two pipelines, $5.3$ come
from the change of point estimator, $1.3$ from the correction of
the residuals and $3.2$ from budgeting the second phase (paired
Monte Carlo standard errors $1.1$, $0.4$ and $0.6$). Adding the
second phase through $\hat{V}_2^{\mathrm{rep}}$ raises coverage by
$3.6$ and $3.2$ points in the linear and nonlinear populations,
where the second-phase share is $23$--$29\%$ and about $17\%$, and
by $0.6$ points in the weak-signal population, where it is $4\%$
(paired Monte Carlo standard errors at most $0.6$ points). What
ignoring the second phase costs in coverage follows the size of the
second phase.

The two proposed estimators separate cleanly along the
adaptive-versus-conditional divide identified above. The analytic
estimator $\hat{V}_1^{\mathrm{loo}}+\hat{V}_2$ gives the
second-phase variance in closed form when the cell structure is
fixed or $s$-measurable, and with trees refitted on each $s_1$ it
measures where that regime ends. In the nonlinear population, where
the tree genuinely adapts, its coverage falls as the training
fraction grows, from $91.2\%$ at $f_1=0.5$ to $89.5\%$ and
$86.7\%$ at $f_1=0.9$, where the held-out part of $s$ is smallest;
in the linear and weak-signal populations it stays within
$91.8$--$93.7\%$ and $93.1$--$93.8\%$, with no trend. Its
second-phase term adds almost nothing to $\hat{V}_1^{\mathrm{loo}}$
alone --- $0.1$, $0.5$ and $0.0$ points of coverage at $f_1=0.7$,
and at most $0.2$ at the other training fractions --- because it
carries only the variability of the cell means; the gap between it
and the replication estimator is thus a second measurement, by a
different instrument, of the structure randomness that
Section~\ref{sec:sim-share} isolates through $\hat{Y}^{(B)}$. The
analytic estimator remains a lower bound on the second-phase cost
and, when the structure is fixed or $s$-measurable, the exact
answer at no refitting cost; with adaptively refitted trees the
estimator to report is the replication one. The
replication estimator
$\hat{V}_1^{\mathrm{loo}}+\hat{V}_2^{\mathrm{rep}}$, which by
Proposition~\ref{prop:rep-unbiased} is conditionally unbiased for the
full second-phase variance \emph{including} structure randomness, is
close to unbiased for the total in all three populations (relative
biases $+6.0\%$, $+8.1\%$ and $-4.2\%$) and brings coverage to
$95.3\%$ and $94.4\%$ in the linear and weak-signal populations ---
inside the $93$--$96\%$ band --- and to $92.2\%$ in the nonlinear
population. Biases of this size are modest on the scale that
matters for intervals: a relative bias $\delta$ in the variance
moves the half-width by about $\delta/2$, so the largest,
$+8.1\%$, lengthens the interval by about $4\%$, and two of the
three are in the conservative direction. Because
$\hat{V}_2^{\mathrm{rep}}$ is exactly unbiased for its target
(Proposition~\ref{prop:rep-unbiased}), these biases are those of
the first-phase term, and the decomposition makes them measurable
directly rather than by subtraction: taking
$V_1=V(\hat{Y}^{(50)})-E_p(V_2)/50$ with $E_p(V_2)$ estimated by
the replicate average, $\hat{V}_1^{\mathrm{loo}}$ overshoots the
true first-phase component by $16.5\%$ and $8.0\%$ in the linear
and nonlinear populations and undershoots it by $4.1\%$ in the
weak-signal one --- the net of the two opposing effects described
in Section~\ref{sec:varest-v1}. The length column should be read in the
same light. In the nonlinear population the replication intervals
are $32\%$ longer than the na\"ive ones ($13.57\%$ against
$10.31\%$ of $Y$), but the Monte Carlo standard deviations of the
two point estimators differ by only $3\%$: almost all of the
difference is the na\"ive variance's underestimation being
corrected, not a cost of the method.

The residual shortfall in the nonlinear population is not a
variance problem, and the Monte Carlo replicates show it directly:
recentering each interval at the Monte Carlo bias raises its
coverage from $92.2\%$ to $94.5\%$, inside the band, whereas the
same recentering leaves the na\"ive interval at $87.8\%$. The
design bias of the point estimator ($-0.88\%$, about one quarter
of its standard error) costs more coverage than its size suggests
because, in this skewed population, the variance estimate moves
with the point estimate (correlation $0.56$): low estimates come
with short intervals, and $7.1\%$ of the intervals miss below the
total against $0.7\%$ above. Recentering at the Monte Carlo bias
is a diagnostic, not a remedy available to a user --- the bias is
unknown in practice. What a user can act on is the bias itself:
a smaller training fraction, which shrinks the bias by
\eqref{eq:composite-bias} at a visible variance cost
(Table~\ref{tab:point}); or, where unbiasedness is worth its
price, the single-split estimator, exactly unbiased at roughly
$1.7$ times the RRMSE at $f_1=0.7$. Further variance inflation is
not the relevant lever.

The random-forest row makes the same diagnosis more sharply: the
variance estimator overshoots ($+14.9\%$) and coverage is still
$90.2\%$, because the forest's design bias ($-1.02\%$, roughly $0.4$
standard errors) is the binding constraint: recentered at that
bias, the same intervals cover $94.2\%$. This is the pattern
anticipated in
Remark~\ref{rem:forests} and documented for bagged estimators since
\citet{wang2014}: for forests the frontier is bias,
not second-phase variance, and we report the result as a finding
about where the method's guarantees stop rather than as a defect of
the variance estimator, whose second-phase component remains exactly
conditionally unbiased by construction.

Two rows of Table~\ref{tab:variance} put the baselines in
context. The cross-fitted estimator with its own out-of-fold
variance covers $95.0$, $92.8$ and $93.8\%$: in the nonlinear
population it sits below the band for the same reason our pipeline
does --- what remains there is bias-limited, not variance-limited
--- and its variance estimator runs $8.7\%$ low where the learner
genuinely adapts, consistent with treating the residual fitting
randomness as negligible. Whether the fold-assignment component of
its variance is material at these sample sizes is measurable with
Theorem~\ref{thm:decomp} --- the fold assignment is a $q$ --- and
is left to the separate comparison study. The node-rule row is the
other baseline: grown under $k(n)=n^{11/20}$, the full-fit
pipeline's na\"ive interval covers $92.0\%$ in the nonlinear
population, against $82.4\%$ at \texttt{rpart} defaults. Most of
the na\"ive failure at defaults is thus the growth condition of
\citet{mcconville2019} being violated, not an intrinsic defect of
the full-fit construction; the reading we take is that the
partition replaces a condition that the analyst must enforce and
verify --- and that default settings silently violate --- with a
design guarantee that holds for any algorithm.

Under the stratified first phase the pattern of
Table~\ref{tab:variance} persists: at $f_1=0.7$ the second-phase
share is $25.5$, $17.5$ and $3.1\%$, coverage with the replication
estimator is $96.8$, $92.6$ and $93.8\%$ (conservative in the
linear population, where the combined estimator overshoots by
$22\%$), $\hat{V}_1^{\mathrm{loo}}$ alone covers $95.2$, $88.7$
and $93.3\%$, and the cross-fitted pipeline covers $95.1$, $92.0$
and $94.0\%$.

For the choice left open in Section~\ref{sec:varest-v2-rep}, the
evidence in these populations points one way: with adaptively fitted
trees we would report the replication estimator, paired as here with a
first-phase component suited to the learner, and reserve the analytic
estimator for cell structures that are fixed or $s$-measurable ---
where Proposition~\ref{prop:v2-known} makes it exact and free.

Because $\hat{Y}_{\mathrm{tree}}$ is the estimator of
\citet{mcconville2019}, the comparison between the two
constructions is direct rather than hypothetical, and in this
configuration they fail in different places. In the nonlinear
population the full-fit estimator carries the larger design bias
($-1.24\%$, against $-0.88\%$ for $\hat{Y}_{TS}^{\mathrm{tree}}$
at $f_1=0.7$), and the variance computed from its in-sample
residuals falls $33.7\%$ short, so that its interval covers
$82.4\%$ of the time. With the partition and the replication
estimator the variance is overestimated by $8.1\%$, and what
remains of the shortfall is the point bias. What the partition
leaves open is theoretical --- the order of $E_p(V_2)$ when the
structure is refitted (Remark~\ref{rem:adaptive}) --- and it does
not prevent that term from being estimated; what in-sample
residuals leave, in this configuration, is an interval that does
not cover what it states. The difference is one of mechanism. In
the full-fit construction the effect of fitting the cells on the
same sample that enters the correction is treated as asymptotically
negligible under a condition on the tree, a minimum node size
growing faster than $n^{1/2}$ \citep[Sec.~2]{mcconville2019}; here
it is removed by the design, and its price appears as an explicit
second-phase term. The distinction matters where such conditions
are not met or are hard to verify: trees grown at default settings,
whose minimum node size does not grow with $n$, more flexible
learners, and forests.

\section{Discussion}
\label{sec:discussion}

This paper set out to provide the design-based uncertainty statement
that accompanies the estimator a practitioner actually reports after
fitting a flexible model on one training subsample. The route was a
two-phase representation whose intermediate term carries all
second-phase randomness, and it delivered three exact,
algorithm-free results --- the collapse identity, the equality of
bias between the single-partition estimator and its
partition-averaged version, and the variance family
$V_1+E_p(V_2)/B$ --- together with computable second-phase variance
formulas for tree-type predictors and two variance estimators that
leave the reported estimate untouched. The simulations support the
package where the theory says they should, and are instructive where
it says nothing: coverage lands in the target band for the linear
and weak-signal populations, the analytic estimator is exact ---
and the share follows its theoretical order across two sample sizes
--- in the $s$-measurable configuration, and the empirical
second-phase share
--- about $17\%$ of total variance in the population where the
tree is the appropriate tool, and from $4\%$ to $23$--$29\%$
across the three, with adaptively refitted trees --- shows that the
component a fixed-prediction analysis omits is far from negligible
in practice; ignoring it costs between $0.6$ and $3.6$ points of
coverage, in step with its size.

The most useful empirical lesson concerns adaptivity. With the tree
structure refitted on each training subsample, the second-phase
variance is dominated by structure randomness, which the
conditional-on-cells formula of Theorem~\ref{thm:trees} does not
carry; the replication estimator, exactly conditionally unbiased for
the full second-phase variance by
Proposition~\ref{prop:rep-unbiased}, absorbs it at the cost of $A$
refits. The division of labor that emerges is clean: analytic and
free when the structure is fixed, replication when it is not.

For practice, the results sort into three statements. If
a flexible learner is needed and a single fit is to be reported,
the partition is what makes a design-based variance statement for
that number possible, and the price of the partition can be
estimated for any algorithm
(Proposition~\ref{prop:rep-unbiased}). If the cell structure is
fixed or $s$-measurable, the analytic estimator gives that price in
closed form and at no refitting cost; if the structure is refitted
adaptively, the replication estimator is the one to report. And
whoever splits a sample and reports a number that depends on the
split carries this second phase in the uncertainty of that number,
whether or not it is budgeted.

The limitations are stated, not implied. We prove no central limit
theorem, and the intervals rest on a working normal approximation
whose adequacy was checked only empirically. The theory for trees gives the second-phase variance in closed form
when the cell structure is fixed or $s$-measurable; the fully
adaptive case is covered by the exact decomposition and the
replication estimator, but the order of $E_p(V_2)$ when the
structure is refitted on each training subsample is open. The
natural conjecture is that under a minimum-node-size condition of
the type used by \citet{mcconville2019} the randomness of the
structure acts on the same cells-of-size-$n/H$ scale and preserves
the order $\{(1-f_1)/f_1\}(H/n)$ of
Corollary~\ref{cor:relative-cost}. The simulations now measure
the two sides of the divide at one sample size: in the
$s$-measurable configuration the share follows the $H/n$ order at
$n=500$ and $n=2{,}000$ (Section~\ref{sec:sim-smedible}), while
with the structure refitted on each $s_1$ it is an order of
magnitude larger (Section~\ref{sec:sim-share}) --- a direct
measurement of structure randomness. What they do not settle is
how the adaptive share scales with $n$, with or without a growing
node-size condition: the node-rule runs fix $k(n)$ at a single
$n$. Settling that order is the open problem this paper leaves. Random forests enter only empirically, and there the
binding constraint is design bias rather than second-phase variance
--- consistent with the long-standing open problem of
\citet{wang2014} --- so bias reduction, not variance refinement, is
the relevant frontier. The executed first-phase designs are
equal-probability --- simple random, and stratified with
proportional allocation, under which the results carry over
(Tables~\ref{tab:stratified} and the closing rows of
Section~\ref{sec:sim-variance}); the estimators and both variance
estimators are written for general $\pi_k$, but their behavior
under unequal-probability ($\pi$ps) first phases remains to be
evaluated, and there the composite identity requires
$\pi$-weighted cell means. The second-phase design is a
separate axis, and one the analyst controls: $q$ is chosen rather
than given. Theorem~\ref{thm:decomp} holds for any $q$, so the
decomposition is unaffected, but the closed form of
Theorem~\ref{thm:trees} and the order of
Corollary~\ref{cor:relative-cost} are derived under
$\mathrm{SRSWOR}$ and would have to be rederived for a stratified
or unequal-probability training subsample. Designing $q$, rather
than taking it as given, is an open direction that the two-phase
formulation makes available. Finally, the choice of the
training fraction was treated as an experimental variable, not
optimized; the exact variance family gives the objective function
such an optimization would minimize.

Beyond these direct extensions, a systematic comparison with the
exactly $pq$-unbiased construction of \citet{sanguiao2021} and
\citet{zhang2026}, with the U/V-statistic variance of
\citet{dharamshi2025}, and with cross-fitted inference at matched
computational budgets --- including the fold-assignment share of
its variance, which Theorem~\ref{thm:decomp} makes measurable ---
is the natural next study; the family formula
of Theorem~\ref{thm:decomp} provides the common yardstick --- the
same decomposition prices every member --- and the equivalence
between prediction-powered inference and the difference estimator
\citep{mozer2026} suggests that the single-partition variance
question, and the answer proposed here, travel beyond the survey
setting in which they were posed.

\appendix
\section{Proofs}
\label{app:proofs}

\subsection{Preliminaries: facts used repeatedly}
\label{app:prelim}

The arguments below are elementary but repetitive, and they rest on
three facts that we record once, in the exact form in which they are
used, so that each proof can be read as a short chain of citations to
them. Throughout, $g$ denotes a function not depending on the
samples, and all integrability requirements are met because $U$ is
finite and, where invoked, (A2) holds.

\medskip
\noindent\textbf{(P1)} \emph{(Centred design weights.)} Since
$E_p(I_k)=\pi_k$,
\[
E_p\Bigl(1-\frac{I_k}{\pi_k}\Bigr)=0
\quad (k\in U),
\qquad\text{hence}\qquad
E_p\Bigl\{\sum_{k\in U}\Bigl(1-\frac{I_k}{\pi_k}\Bigr)
g(\mathbf{x}_k)\Bigr\}=0
\]
for every fixed $g$. The restriction to fixed $g$ is essential: the
identity fails as soon as the function multiplying $1-I_k/\pi_k$
depends on $s$, and that failure is precisely the source of the
design bias computed in Proposition~\ref{prop:bias}. We also record
$E_p\{(1-I_k/\pi_k)^2\}=V_p(I_k)/\pi_k^2=(1-\pi_k)/\pi_k\leq
1/\pi_k$. Because the factor is centred, for any integrable
statistic $B$ of $s$ we have
$E_p\{(1-I_k/\pi_k)B\}=\mathrm{Cov}_p(1-I_k/\pi_k,\,B)$, and
likewise for $N-\hat{N}_\pi=\sum_{k\in U}(1-I_k/\pi_k)$.

\medskip
\noindent\textbf{(P2)} \emph{(Mean of an $\mathrm{SRSWOR}$ sample
from a finite set.)} Let $d$ be an $\mathrm{SRSWOR}$ sample of size
$m$ drawn from a finite set $D$ of size $M\geq m\geq 1$, and let
$\bar{y}_d=m^{-1}\sum_{k\in d}y_k$. Then
\[
E(\bar{y}_d)=\bar{y}_D,
\qquad
V(\bar{y}_d)=\Bigl(\frac{1}{m}-\frac{1}{M}\Bigr)S^2_{y,D},
\qquad
S^2_{y,D}=\frac{1}{M-1}\sum_{k\in D}(y_k-\bar{y}_D)^2 .
\]
Every second-phase computation in this paper is an instance of (P2)
with $D=s$ (so $M=n$) and $d=s_1$ (so $m=n_1$), or with $D$ and $d$
the parts of $s$ and $s_1$ falling in a single cell: the role of the
textbook population is played by the realized first-phase sample
(the reversal of Section~\ref{sec:setup-classical}). Under (A3),
$1/n_1-1/n\leq(1-f_1)/(f_1n)$ by \eqref{eq:rounding}, with
equality if and only if $f_1n\in\mathbb{N}$.

\medskip
\noindent\textbf{(P3)} \emph{($s$-measurability as a computing
rule.)} If $W$ is $s$-measurable and $Z$ is any statistic of
$(s,s_1)$, then conditionally on $s$ the quantity $W$ behaves as a
constant:
\[
E_q(W\mid s)=W,
\qquad
V_q(W\mid s)=0,
\qquad
E_q(WZ\mid s)=W\,E_q(Z\mid s),
\]
\[
V_q(W+Z\mid s)=V_q(Z\mid s),
\qquad
V_q(WZ\mid s)=W^2\,V_q(Z\mid s).
\]
The quantities $I_k$, $\pi_k$, $\Delta_{k\ell}$, $\hat{m}$,
$\hat{Y}_\pi$, $\hat{N}_\pi$, $\bar{y}_s$, $S^2_{y,s}$, and in the
tree case $N_h$, $n_h$, $D_h$, $S^2_{y,s,h}$ and the ($s$-measurable)
cell system itself, are all $s$-measurable; the training fit
$\hat{m}_1$ and the training cell counts $n_{1h}$ are not. Almost
every conditional computation below consists of sorting the terms of
an expression into these two classes and applying (P3) to the first.

\medskip
\noindent\textbf{(P4)} \emph{(Nested $\mathrm{SRSWOR}$: marginal
and conditional laws.)} If $s$ is $\mathrm{SRSWOR}(n)$ from $U$
and, given $s$, $s_1$ is $\mathrm{SRSWOR}(n_1)$ from $s$, then
marginally $s_1$ is an $\mathrm{SRSWOR}(n_1)$ sample from $U$, and
conditionally on $s_1$ the complement $s_2=s\setminus s_1$ is an
$\mathrm{SRSWOR}(n-n_1)$ sample from $U\setminus s_1$; in
particular $\Pr(k\in s_2\mid s_1)=(n-n_1)/(N-n_1)$ for
$k\in U\setminus s_1$. Both claims follow by counting: each
compatible pair $(s,s_1)$ has the same probability, so every
size-$n_1$ subset of $U$ is equally likely to be $s_1$, and, given
$s_1$, every size-$(n-n_1)$ subset of $U\setminus s_1$ is equally
likely to be $s_2$. The same holds within each stratum when both
phases are stratified $\mathrm{SRSWOR}$ over the same strata.

\subsection{Proof of Lemma~\ref{lem:collapse}}
\label{app:collapse}

Let $\hat{g}$ be an arbitrary $s$-measurable prediction rule, placed
in the role of $\hat{m}$ in the three terms of
\eqref{eq:ysl-threeterm}:
\[
T_0(\hat{g})=\sum_{k\in U}\hat{g}(\mathbf{x}_k),
\quad
T_1(\hat{g})=\sum_{k\in U}\Bigl(1-\frac{I_k}{\pi_k}\Bigr)
\{\hat{m}_1(\mathbf{x}_k)-\hat{g}(\mathbf{x}_k)\},
\quad
T_2(\hat{g})=\sum_{k\in s}\frac{y_k-\hat{g}(\mathbf{x}_k)}{\pi_k}.
\]
Split the weight in $T_1(\hat{g})$ and use
$\sum_{k\in U}I_ka_k=\sum_{k\in s}a_k$:
\begin{align}
T_1(\hat{g})
&=\sum_{k\in U}\{\hat{m}_1(\mathbf{x}_k)-\hat{g}(\mathbf{x}_k)\}
-\sum_{k\in U}\frac{I_k}{\pi_k}
\{\hat{m}_1(\mathbf{x}_k)-\hat{g}(\mathbf{x}_k)\}
\nonumber\\
&=\sum_{k\in U}\hat{m}_1(\mathbf{x}_k)
-\sum_{k\in U}\hat{g}(\mathbf{x}_k)
-\sum_{k\in s}\frac{\hat{m}_1(\mathbf{x}_k)}{\pi_k}
+\sum_{k\in s}\frac{\hat{g}(\mathbf{x}_k)}{\pi_k}.
\label{eq:collapse-step}
\end{align}
Adding $T_0(\hat{g})$ cancels the second term of
\eqref{eq:collapse-step}; adding
$T_2(\hat{g})=\sum_{k\in s}y_k/\pi_k
-\sum_{k\in s}\hat{g}(\mathbf{x}_k)/\pi_k$ cancels the fourth. What
remains is
\[
T_0(\hat{g})+T_1(\hat{g})+T_2(\hat{g})
=\sum_{k\in U}\hat{m}_1(\mathbf{x}_k)
+\sum_{k\in s}\frac{y_k-\hat{m}_1(\mathbf{x}_k)}{\pi_k},
\]
which is \eqref{eq:ysl-collapsed}. Three consequences are worth
isolating, since all three are used later.

First, the right-hand side does not involve $\hat{g}$: the
three-term representation is invariant to the choice of
$s$-measurable reference rule. Taking $\hat{g}=\hat{m}$ recovers
\eqref{eq:ysl-threeterm}, so the full-sample fit never has to be
computed and serves only to interpret the decomposition.

Second --- this is the form in which the lemma is invoked in
Appendix~\ref{app:rb-identity} --- the calculation used no property
of $\hat{m}_1$ beyond its occupying two particular slots, and it is
\emph{linear} in whatever occupies them. Hence, for an arbitrary
function $\psi$,
\begin{equation}
\label{eq:collapse-general}
\sum_{k\in U}\hat{g}(\mathbf{x}_k)
+\sum_{k\in U}\Bigl(1-\frac{I_k}{\pi_k}\Bigr)
\{\psi(\mathbf{x}_k)-\hat{g}(\mathbf{x}_k)\}
+\sum_{k\in s}\frac{y_k-\hat{g}(\mathbf{x}_k)}{\pi_k}
=\sum_{k\in U}\psi(\mathbf{x}_k)
+\sum_{k\in s}\frac{y_k-\psi(\mathbf{x}_k)}{\pi_k}.
\end{equation}
Nothing requires $\psi$ to be a fitted rule, or even to be random.

Third, with $\hat{g}=\hat{m}$ both $T_0$ and $T_2$ are functions of
$s$ and $\hat{m}$ alone, hence $s$-measurable, and
$\hat{Y}_{TS}-(T_0+T_2)=T_1$; all dependence on $s_1$ is therefore
carried by $T_1$. \qed

\subsection{Proof of Proposition~\ref{prop:degenerate}}
\label{app:degenerate}

\emph{The closed form.} With $\hat{m}_1(\mathbf{x})\equiv
\bar{y}_{s_1}$ the collapsed representation
\eqref{eq:ysl-collapsed} has a constant prediction rule, so its two
terms are
\[
\sum_{k\in U}\hat{m}_1(\mathbf{x}_k)=N\bar{y}_{s_1},
\qquad
\sum_{k\in s}\frac{y_k-\bar{y}_{s_1}}{\pi_k}
=\sum_{k\in s}\frac{y_k}{\pi_k}
-\bar{y}_{s_1}\sum_{k\in s}\frac{1}{\pi_k}
=\hat{Y}_\pi-\hat{N}_\pi\bar{y}_{s_1},
\]
and adding them gives
$\hat{Y}_{TS}=N\bar{y}_{s_1}+\hat{Y}_\pi-\hat{N}_\pi\bar{y}_{s_1}
=\hat{Y}_\pi+(N-\hat{N}_\pi)\bar{y}_{s_1}$, which is
\eqref{eq:degenerate}. The training subsample enters through the
single scalar $\bar{y}_{s_1}$, multiplied by the first-phase level
error $N-\hat{N}_\pi$.

\emph{(a)} Under $\mathrm{SRSWOR}(n)$ at the first phase
$\pi_k=n/N$ for every $k$, so $\hat{N}_\pi=\sum_{k\in s}N/n=N$ and
the second term of \eqref{eq:degenerate} is identically zero --- not
merely zero in expectation. Hence $\hat{Y}_{TS}=\hat{Y}_\pi
=N\bar{y}_s$ whatever subsample is drawn: the partition has no
effect on the realized estimate.

\emph{(b)} Under (A3), $s_1$ is an $\mathrm{SRSWOR}(n_1)$ sample from
the finite set $s$, so $E_q(\bar{y}_{s_1}\mid s)=\bar{y}_s$ by (P2).
Since $\hat{Y}_\pi$ and $N-\hat{N}_\pi$ are $s$-measurable, (P3)
gives
\[
E_q(\hat{Y}_{TS}\mid s)
=\hat{Y}_\pi+(N-\hat{N}_\pi)\,E_q(\bar{y}_{s_1}\mid s)
=\hat{Y}_\pi+(N-\hat{N}_\pi)\,\bar{y}_s .
\]
The right-hand side is exactly $\hat{Y}_{RB}$ for this algorithm,
consistently with Proposition~\ref{prop:rb-identity}, and the second
phase contributes no conditional bias.

\emph{(c)} By (P3), conditionally on $s$ the only random quantity in
\eqref{eq:degenerate} is $\bar{y}_{s_1}$, multiplied by the
$s$-measurable constant $N-\hat{N}_\pi$; hence
\[
V_q(\hat{Y}_{TS}\mid s)
=(N-\hat{N}_\pi)^2\,V_q(\bar{y}_{s_1}\mid s)
=(N-\hat{N}_\pi)^2\Bigl(\frac{1}{n_1}-\frac{1}{n}\Bigr)S^2_{y,s},
\]
by (P2) with $D=s$, $M=n$, $m=n_1$. This is
\eqref{eq:degenerate-v2}, valid exactly for every $n$ and $n_1$; the
bound \eqref{eq:degenerate-v2-f} then follows from
\eqref{eq:rounding}, with equality when $f_1n\in\mathbb{N}$. \qed

\begin{remark}[$pq$-unbiasedness in the degenerate case]
\label{rem:degenerate-bias}
Part (b) concerns the second phase alone. Taking $E_p$ in
\eqref{eq:degenerate} and using $E_p(\hat{Y}_\pi)=Y$ and
$E_p(\hat{N}_\pi)=N$, the factor $N-\hat{N}_\pi$ is centred, so by (P1) the
expectation is a covariance and gives the exact design bias
\[
E(\hat{Y}_{TS})-Y
=E_p\{(N-\hat{N}_\pi)\bar{y}_s\}
=\mathrm{Cov}_p(N-\hat{N}_\pi,\bar{y}_s)
=-\mathrm{Cov}_p(\hat{N}_\pi,\bar{y}_s),
\]
which is \eqref{eq:bias-exact} specialized to
$\bar{m}_s\equiv\bar{y}_s$. It vanishes whenever $\hat{N}_\pi$ is
design-constant, in particular under (a). In general, by
Cauchy--Schwarz and (A1)--(A2), $V_p(\hat{N}_\pi)=O(N^2/n)$ and
$V_p(\bar{y}_s)=O(1/n)$, so the bias is $O(N/n)$ and the relative
bias is $O(n^{-1})$ --- an order smaller than the standard error,
which is $O(n^{-1/2})$. The degenerate case is thus exactly
$pq$-unbiased under equal probabilities and approximately so in
general, with no appeal to (A4$'$).
\end{remark}

\subsection{Proof of Proposition~\ref{prop:composite} and
Corollary~\ref{cor:fullfit-plugin}}
\label{app:composite}

Write $e_k=y_k-\hat{m}_1(\mathbf{x}_k)$ for $k\in U$, well defined
given $s_1$. Because $\hat{m}_1$ predicts the $s_1$-cell mean,
$\sum_{k\in s_{1h}}e_k=0$ for every cell, hence
$\sum_{k\in s_1}e_k=0$ and
$\sum_{k\in s}e_k=\sum_{k\in s_2}e_k$. With $\pi_k\equiv n/N$,
the collapsed form \eqref{eq:ysl-collapsed} gives
\[
\hat{Y}_{TS}
=\hat{Y}_{P_1}+\frac{N}{n}\sum_{k\in s}e_k
=\hat{Y}_{P_1}+\frac{N}{n}\sum_{k\in s_2}e_k .
\]
For \eqref{eq:sanguiao}, adding and subtracting
$\hat{m}_1(\mathbf{x}_k)$ on $s_1$,
\[
\hat{Y}_1^{M}
=\sum_{k\in s_1}y_k
+\sum_{k\in U\setminus s_1}\hat{m}_1(\mathbf{x}_k)
+\frac{1}{\pi_2}\sum_{k\in s_2}e_k
=\hat{Y}_{P_1}+\sum_{k\in s_1}e_k
+\frac{1}{\pi_2}\sum_{k\in s_2}e_k
=\hat{Y}_{P_1}+\frac{1}{\pi_2}\sum_{k\in s_2}e_k .
\]
Hence
$\hat{Y}_{TS}-\hat{Y}_{P_1}
=(N/n)\,\pi_2\,\{\hat{Y}_1^{M}-\hat{Y}_{P_1}\}
=c\,\{\hat{Y}_1^{M}-\hat{Y}_{P_1}\}$, which is
\eqref{eq:composite}. Under $\mathrm{SRSWOR}$ at both phases, (P4)
gives $\pi_2=(n-n_1)/(N-n_1)$, so, with $f_1=n_1/n$ the realized
fraction,
\[
c=\frac{N}{n}\cdot\frac{n-n_1}{N-n_1}
=\frac{1-f_1}{1-f_1\,n/N}\,,
\]
and $c<1$ is equivalent to $N(n-n_1)<n(N-n_1)$, that is, to
$n_1(N-n)>0$. Under stratified $\mathrm{SRSWOR}$ at both phases
over the same strata with proportional allocation and
$n_{1g}=f_1n_g$, (P4) applied within each stratum gives
$\Pr(k\in s_2\mid s_1)=(n_g-n_{1g})/(N_g-n_{1g})$ for
$k\in U_g\setminus s_1$, the same constant across strata because
$n_g/N_g=n/N$, and the display above applies with the corrections
summed within strata.

For \eqref{eq:composite-bias}: conditionally on $s_1$, using
$\Pr(k\in s_2\mid s_1)=\pi_2$,
\[
E\Bigl(\frac{1}{\pi_2}\sum_{k\in s_2}e_k\,\Big|\,s_1\Bigr)
=\sum_{k\in U\setminus s_1}e_k ,
\]
so
$E(\hat{Y}_1^{M}\mid s_1)
=\sum_{k\in s_1}y_k
+\sum_{k\in U\setminus s_1}\hat{m}_1(\mathbf{x}_k)
+\sum_{k\in U\setminus s_1}\{y_k-\hat{m}_1(\mathbf{x}_k)\}=Y$:
the single-split estimator is exactly $pq$-unbiased for any rule
fitted on $s_1$ \citep[Prop.~6]{sanguiao2021}. Taking expectations
in \eqref{eq:composite} yields \eqref{eq:composite-bias}.

For Corollary~\ref{cor:fullfit-plugin}, apply the vanishing-sum
fact with $s_1=s$: a cell-mean rule fitted on all of $s$ has
$\sum_{k\in s_h}\{y_k-\hat{m}(\mathbf{x}_k)\}=0$ in every
cell, so the correction term
$(N/n)\sum_{k\in s}\{y_k-\hat{m}(\mathbf{x}_k)\}$ vanishes
identically. \qed

\subsection{Proof of Proposition~\ref{prop:rb-identity}}
\label{app:rb-identity}

The proof has three steps: identify what is random given $s$; take
the conditional expectation term by term; and recognize the result
as an instance of the collapse identity.

\emph{Step 1: what is random given $s$.} Write $\hat{Y}_{TS}$ in the
three-term form \eqref{eq:ysl-threeterm}, with reference rule
$\hat{m}$. By Lemma~\ref{lem:collapse} the terms
$T_0=\sum_{k\in U}\hat{m}(\mathbf{x}_k)$ and
$T_2=\sum_{k\in s}\{y_k-\hat{m}(\mathbf{x}_k)\}/\pi_k$ are
$s$-measurable, and so are the weights $1-I_k/\pi_k$ occurring in
$T_1$, since $I_k$ and $\pi_k$ refer to the first phase only.
Conditionally on $s$, therefore, the entire second-phase randomness
of the expression sits in the training fit $\hat{m}_1$, which
depends on which subset $s_1$ was drawn. By (P3),
\[
E_q(T_0\mid s)=T_0,
\qquad
E_q(T_2\mid s)=T_2 .
\]

\emph{Step 2: conditional expectation of the intermediate term.} The
sum defining $T_1$ is finite and each summand is integrable, by the
hypothesis $E_q\{|\hat{m}_1(\mathbf{x}_k)|\mid s\}<\infty$; hence
the expectation may be taken inside the sum, and the $s$-measurable
factors may be taken outside each expectation by (P3):
\begin{align*}
E_q(T_1\mid s)
&=\sum_{k\in U}\Bigl(1-\frac{I_k}{\pi_k}\Bigr)
E_q\{\hat{m}_1(\mathbf{x}_k)-\hat{m}(\mathbf{x}_k)\mid s\}\\[2pt]
&=\sum_{k\in U}\Bigl(1-\frac{I_k}{\pi_k}\Bigr)
\{\bar{m}_s(\mathbf{x}_k)-\hat{m}(\mathbf{x}_k)\},
\end{align*}
by the definition \eqref{eq:rb-rule} of $\bar{m}_s$. Only linearity
of expectation over a finite sum has been used. The effect of the
operation is purely local: at each $\mathbf{x}_k$ the prediction of
the realized training fit has been replaced by its average over the
draws of $s_1$, and the surrounding design structure is untouched.

\emph{Step 3: recognizing the collapse.} Adding the three
conditional expectations,
\[
E_q(\hat{Y}_{TS}\mid s)
=\sum_{k\in U}\hat{m}(\mathbf{x}_k)
+\sum_{k\in U}\Bigl(1-\frac{I_k}{\pi_k}\Bigr)
\{\bar{m}_s(\mathbf{x}_k)-\hat{m}(\mathbf{x}_k)\}
+\sum_{k\in s}\frac{y_k-\hat{m}(\mathbf{x}_k)}{\pi_k},
\]
which is exactly the left-hand side of the algebraic identity
\eqref{eq:collapse-general} with $\hat{g}=\hat{m}$ and
$\psi=\bar{m}_s$. That identity holds for \emph{any} function in the
slot of $\psi$ --- the collapse being linear in the rule placed
there, and requiring nothing of it --- so it applies verbatim, and
its right-hand side gives
\[
E_q(\hat{Y}_{TS}\mid s)
=\sum_{k\in U}\bar{m}_s(\mathbf{x}_k)
+\sum_{k\in s}\frac{y_k-\bar{m}_s(\mathbf{x}_k)}{\pi_k}
=\hat{Y}_{RB},
\]
which is \eqref{eq:rb-estimator}. In words: partition averaging,
which is an expectation, and the collapse, which is a linear
algebraic identity, commute. Averaging the fitted rule over
partitions and then forming the model-assisted estimator yields the
same object as forming the estimator and then averaging it --- and
this is why $\hat{Y}_{RB}$ is again a model-assisted estimator, of
the standard form, built on the $s$-measurable rule $\bar{m}_s$.

\emph{Consequence (a).} Taking $E_p$ on both sides and using $E=E_pE_q$,
\[
E(\hat{Y}_{TS})=E_p\{E_q(\hat{Y}_{TS}\mid s)\}
=E_p(\hat{Y}_{RB})=E(\hat{Y}_{RB}),
\]
the last equality because $\hat{Y}_{RB}$ is $s$-measurable and so
does not change under $E_q(\cdot\mid s)$. The two estimators have
identical design bias, exactly, with no condition beyond
integrability, and in particular for every algorithm, every pair
$(p,q)$ and every sample size.

\emph{Consequence (b).} For
$\hat{Y}^{(B)}=B^{-1}\sum_{b=1}^{B}\hat{Y}_{TS}^{(b)}$ with
$s_1^{(1)},\ldots,s_1^{(B)}$ drawn independently from
$q(\cdot\mid s)$, each replicate satisfies
$E_q(\hat{Y}_{TS}^{(b)}\mid s)=\hat{Y}_{RB}$ by the above; hence, by
linearity, $E_q(\hat{Y}^{(B)}\mid s)=B^{-1}\cdot B\cdot\hat{Y}_{RB}
=\hat{Y}_{RB}$ for every $B$, the case $B=1$ included. Averaging
therefore changes the variance but never the conditional mean, which
is the precise sense in which partition averaging here is a
variance device and nothing else. \qed

\subsection{Proof of Proposition~\ref{prop:bias}}
\label{app:bias}

\emph{(a)} By Proposition~\ref{prop:rb-identity} the bias of
$\hat{Y}_{TS}$ equals that of $\hat{Y}_{RB}$, so it suffices to
compute $E_p(\hat{Y}_{RB})-Y$. Rewrite $\hat{Y}_{RB}$ by
introducing the Horvitz--Thompson estimator, using
$\sum_{k\in s}a_k=\sum_{k\in U}I_ka_k$:
\[
\hat{Y}_{RB}
=\sum_{k\in U}\bar{m}_s(\mathbf{x}_k)
+\sum_{k\in s}\frac{y_k}{\pi_k}
-\sum_{k\in s}\frac{\bar{m}_s(\mathbf{x}_k)}{\pi_k}
=\hat{Y}_\pi
+\sum_{k\in U}\Bigl(1-\frac{I_k}{\pi_k}\Bigr)
\bar{m}_s(\mathbf{x}_k).
\]
Since $E_p(\hat{Y}_\pi)=Y$, taking $E_p$ leaves only the second
term:
\[
E(\hat{Y}_{TS})-Y
=E_p\Bigl\{\sum_{k\in U}\Bigl(1-\frac{I_k}{\pi_k}\Bigr)
\bar{m}_s(\mathbf{x}_k)\Bigr\}
=\sum_{k\in U}E_p\Bigl\{\Bigl(1-\frac{I_k}{\pi_k}\Bigr)
\bar{m}_s(\mathbf{x}_k)\Bigr\}.
\]
Each factor $1-I_k/\pi_k$ is centred, so by (P1) each term is a
covariance with $B=\bar{m}_s(\mathbf{x}_k)$:
\[
E(\hat{Y}_{TS})-Y
=\sum_{k\in U}\mathrm{Cov}_p\Bigl(1-\frac{I_k}{\pi_k},
\bar{m}_s(\mathbf{x}_k)\Bigr)
=-\sum_{k\in U}\mathrm{Cov}_p\Bigl(\frac{I_k}{\pi_k},
\bar{m}_s(\mathbf{x}_k)\Bigr),
\]
which is \eqref{eq:bias-exact}. The identity is exact and makes the
mechanism visible: bias arises \emph{only} through the covariation
between sample membership and the rule fitted from the sample, and
would vanish identically, by (P1), for any rule not depending on
$s$.

\emph{(b)} Let $m_\nu$ be the deterministic sequence supplied by
(A4) and put
\[
a_k=1-\frac{I_k}{\pi_k},
\qquad
d_k=\bar{m}_s(\mathbf{x}_k)-m_\nu(\mathbf{x}_k).
\]
Because $m_\nu$ is non-random, (P1) gives
$E_p\{\sum_{k\in U}a_k m_\nu(\mathbf{x}_k)\}=0$; subtracting this
from part (a) leaves
$E(\hat{Y}_{TS})-Y=E\{\sum_{k\in U}a_kd_k\}$. Apply the
Cauchy--Schwarz inequality twice --- first to the sum over $k$, then
to the expectation:
\[
\Bigl|E\Bigl(\sum_{k\in U}a_kd_k\Bigr)\Bigr|
\leq E\Bigl\{\Bigl(\sum_{k\in U}a_k^2\Bigr)^{1/2}
\Bigl(\sum_{k\in U}d_k^2\Bigr)^{1/2}\Bigr\}
\leq\Bigl\{E_p\Bigl(\sum_{k\in U}a_k^2\Bigr)\Bigr\}^{1/2}
\Bigl\{E\Bigl(\sum_{k\in U}d_k^2\Bigr)\Bigr\}^{1/2}.
\]
For the first factor, (P1) and (A1) give
$E_p(a_k^2)=(1-\pi_k)/\pi_k\leq 1/\pi_k\leq N/(\lambda n)$, so
$E_p(\sum_Ua_k^2)\leq N^2/(\lambda n)$. The second factor equals
$N\delta_n^2$, by the definition of $\delta_n^2$ in (A4). Hence
\[
\bigl|E(\hat{Y}_{TS})-Y\bigr|
\leq\Bigl(\frac{N^2}{\lambda n}\Bigr)^{1/2}
\bigl(N\delta_n^2\bigr)^{1/2}
=N\Bigl(\frac{N}{\lambda n}\Bigr)^{1/2}\delta_n,
\]
and dividing by $N$ gives the stated bound. The bound is driven
entirely by $\delta_n$, a property of the algorithm: the $y$-values
enter only through the boundedness assumed in (A2). \qed

\subsection{Proof of Theorem~\ref{thm:decomp}}
\label{app:decomp}

\emph{The case $B=1$.} By Lemma~\ref{lem:collapse},
$\hat{Y}_{TS}=(T_0+T_2)+T_1$ with $T_0+T_2$ $s$-measurable, so by
(P3) that part contributes nothing to the conditional variance:
\[
V_q(\hat{Y}_{TS}\mid s)=V_q(T_1\mid s)=V_2 .
\]
By Proposition~\ref{prop:rb-identity},
$E_q(\hat{Y}_{TS}\mid s)=\hat{Y}_{RB}$, so the first term of \eqref{eq:tp-decomp} is $V_p(\hat{Y}_{RB})=V_1$. Substituting both
into \eqref{eq:tp-decomp},
\[
V(\hat{Y}_{TS})
=V_p\{E_q(\hat{Y}_{TS}\mid s)\}
+E_p\{V_q(\hat{Y}_{TS}\mid s)\}
=V_1+E_p(V_2),
\]
which is \eqref{eq:main-decomp}. No approximation enters at any
point, and no property of the algorithm is used beyond the
integrability required for the conditional moments to exist; in
particular the identity holds for algorithms with internal
randomization, whose variability is then simply part of $V_2$.

\emph{General $B$.} Write
$\hat{Y}^{(B)}=(T_0+T_2)+B^{-1}\sum_{b=1}^{B}T_1^{(b)}$, where
$T_1^{(b)}$ is the intermediate term computed from the $b$th draw
$s_1^{(b)}$. Given $s$, those draws are independent and identically
distributed, hence so are $T_1^{(1)},\ldots,T_1^{(B)}$, with common
conditional mean $E_q(T_1\mid s)$ and common conditional variance
$V_2$. Therefore, using (P3) for the $s$-measurable part and
independence for the average,
\[
E_q(\hat{Y}^{(B)}\mid s)=(T_0+T_2)+E_q(T_1\mid s)=\hat{Y}_{RB},
\qquad
V_q(\hat{Y}^{(B)}\mid s)
=\frac{1}{B^2}\sum_{b=1}^{B}V_q(T_1^{(b)}\mid s)
=\frac{V_2}{B}.
\]
Applying \eqref{eq:tp-decomp} once more gives $V(\hat{Y}^{(B)})=V_1+E_p(V_2)/B$,
which is \eqref{eq:family}. The first component is unchanged because
the conditional mean does not depend on $B$; only the second-phase
component is divided by the number of refits, which is the exact
sense in which the family interpolates between the single-partition
estimator and its partition-averaged limit. \qed

\subsection{Proof of Proposition~\ref{prop:av1}}
\label{app:av1}

Let $\hat{Y}_{\mathrm{dif}}$ be the infeasible difference estimator
of (A4$'$), built from the deterministic limit rule $m_\nu$, and set
$D=\hat{Y}_{RB}-\hat{Y}_{\mathrm{dif}}$, so that
$\hat{Y}_{RB}=\hat{Y}_{\mathrm{dif}}+D$.

Because $m_\nu$ does not depend on the sample,
$\hat{Y}_{\mathrm{dif}}=\sum_{k\in U}m_\nu(\mathbf{x}_k)
+\sum_{k\in U}I_kE_k/\pi_k$ is a linear statistic in the inclusion
indicators with \emph{fixed} coefficients
$E_k=y_k-m_\nu(\mathbf{x}_k)$. Its design variance is therefore the
classical Horvitz--Thompson expression,
\[
V_p(\hat{Y}_{\mathrm{dif}})
=\sum_{k\in U}\sum_{\ell\in U}\Delta_{k\ell}\,
\frac{E_k}{\pi_k}\,\frac{E_\ell}{\pi_\ell},
\]
exactly --- the double sum appearing in \eqref{eq:av1}. Under
(A1)--(A2) it is $O(N^2/n)$: the $E_k$ are uniformly bounded by
(A2); the $N(N-1)$ off-diagonal terms are bounded by $C_1/n$ each in
absolute value after normalization, contributing $O(N^2/n)$; and the
$N$ diagonal terms are bounded by $C/\pi_k\leq CN/(\lambda n)$,
contributing $O(N^2/n)$ as well.

Expanding the variance of a sum,
\[
V_1=V_p(\hat{Y}_{RB})
=V_p(\hat{Y}_{\mathrm{dif}})
+2\,\mathrm{Cov}_p(\hat{Y}_{\mathrm{dif}},D)+V_p(D),
\qquad
|\mathrm{Cov}_p(\hat{Y}_{\mathrm{dif}},D)|
\leq\{V_p(\hat{Y}_{\mathrm{dif}})\,V_p(D)\}^{1/2},
\]
the inequality by Cauchy--Schwarz. Condition (A4$'$) states that
$E[\{N^{-1}(\hat{Y}_{RB}-\hat{Y}_{\mathrm{dif}})\}^2]=o(n^{-1})$,
that is $E_p(D^2)=o(N^2/n)$; hence $V_p(D)\leq E_p(D^2)=o(N^2/n)$
and $|\mathrm{Cov}_p(\hat{Y}_{\mathrm{dif}},D)|
\leq\{O(N^2/n)\cdot o(N^2/n)\}^{1/2}=o(N^2/n)$. Both correction
terms are thus $o(N^2/n)$, which is \eqref{eq:av1}.

The proof isolates exactly where (A4$'$) does its work: it is the
condition allowing the \emph{estimated} rule to be replaced by the
limit rule inside the variance to first order. Nothing in
Sections~\ref{sec:properties-decomp}
and~\ref{sec:properties-trees} --- the second-phase results that
constitute this paper's contribution --- relies on it. \qed

\subsection{Proof of Theorem~\ref{thm:trees}}
\label{app:trees}

\emph{Reduction to cell means.} Under \eqref{eq:ps-rule} the
training fit is constant on each cell, so
\[
T_1'
:=\sum_{k\in U}\Bigl(1-\frac{I_k}{\pi_k}\Bigr)
\hat{m}_1(\mathbf{x}_k)
=\sum_{h=1}^{H}\bar{y}_{s_{1h}}
\sum_{k\in U_h}\Bigl(1-\frac{I_k}{\pi_k}\Bigr)
=\sum_{h=1}^{H}D_h\,\bar{y}_{s_{1h}},
\]
since $\sum_{k\in U_h}(1-I_k/\pi_k)
=N_h-\sum_{k\in s_h}1/\pi_k=N_h-\hat{N}_{\pi,h}=D_h$. The
remaining part of $T_1$, namely
$-\sum_{k\in U}(1-I_k/\pi_k)\hat{m}(\mathbf{x}_k)$, is
$s$-measurable, so $V_q(T_1\mid s)=V_q(T_1'\mid s)$ by (P3). Thus
the second-phase variance is entirely a question about the joint law
of the $H$ training cell means, weighted by the $s$-measurable
coefficients $D_h$.

\emph{(i) Cell-stratified second phase.} If $n_{1h}$ units are drawn
by $\mathrm{SRSWOR}$ within $s_h$, independently across cells,
then the means $\bar{y}_{s_{1h}}$, $h=1,\ldots,H$, are
conditionally independent given $s$, and (P2) applied within cell
$h$ (with $D=s_h$, $M=n_h$, $m=n_{1h}$) gives
$V_q(\bar{y}_{s_{1h}}\mid s)=(1/n_{1h}-1/n_h)S^2_{y,s,h}$.
Hence, by independence and (P3),
\[
V_2=V_q(T_1'\mid s)
=\sum_{h=1}^{H}D_h^2\,V_q(\bar{y}_{s_{1h}}\mid s)
=\sum_{h=1}^{H}D_h^2
\Bigl(\frac{1}{n_{1h}}-\frac{1}{n_h}\Bigr)S^2_{y,s,h},
\]
which is \eqref{eq:v2-trees-exact}. In this scheme $n_{1h}$ is fixed
by design, so every ingredient is $s$-measurable --- the fact
exploited in Proposition~\ref{prop:v2-known}.

\emph{(ii) Global $\mathrm{SRSWOR}$: the conditional variance.} Now
the training cell counts $\mathbf{n}_1=(n_{11},\ldots,n_{1H})$ are
themselves random: $\mathbf{n}_1$ follows a multivariate
hypergeometric law, and each $n_{1h}$ is hypergeometric with mean
$n_1n_h/n$. Conditionally on $\mathbf{n}_1$, however, every
configuration of subsets compatible with those counts is equally
likely, so the conditional law factorizes over cells into
independent $\mathrm{SRSWOR}(n_{1h})$ draws within $s_h$:
drawing $n_1$ units at random from $s$ is the same as first drawing
the vector of counts and then, given the counts, drawing at random
within each cell.

Apply the law of total variance a second time, now given $s$ and the
event $\Omega_0$, using $\mathbf{n}_1$ as the conditioning variable:
\[
V_q(T_1'\mid s,\Omega_0)
=E_q\{V_q(T_1'\mid s,\mathbf{n}_1)\mid s,\Omega_0\}
+V_q\{E_q(T_1'\mid s,\mathbf{n}_1)\mid s,\Omega_0\}.
\]
On $\Omega_0$ every cell satisfies $n_{1h}\geq 1$, so by (P2)
$E_q(\bar{y}_{s_{1h}}\mid s,\mathbf{n}_1)=\bar{y}_{s_h}$,
which does not depend on $\mathbf{n}_1$; hence
$E_q(T_1'\mid s,\mathbf{n}_1)=\sum_hD_h\bar{y}_{s_h}$ is an
$s$-measurable constant and the second term vanishes. The first
term is, by part (i) applied conditionally on $\mathbf{n}_1$,
\[
V_q(T_1'\mid s,\Omega_0)
=E_q\Bigl[\sum_{h=1}^{H}D_h^2
\Bigl(\frac{1}{n_{1h}}-\frac{1}{n_h}\Bigr)S^2_{y,s,h}
\;\Big|\;\Omega_0\Bigr],
\]
which is \eqref{eq:v2-trees-global}. The only change from (i) is
that $1/n_{1h}$ must now be averaged over the random counts.

\emph{(ii) The probability of an empty cell.} The subsamples of size
$n_1$ avoiding $s_h$ entirely number $\binom{n-n_h}{n_1}$, so
\[
\Pr(n_{1h}=0\mid s)
=\frac{\binom{n-n_h}{n_1}}{\binom{n}{n_1}}
=\prod_{j=0}^{n_h-1}\frac{n-n_1-j}{n-j}
\leq\Bigl(1-\frac{n_1}{n}\Bigr)^{n_h}
\leq(1-f_1)^{n_h},
\]
where the first inequality holds because
$(n-n_1-j)/(n-j)=1-n_1/(n-j)$ is decreasing in $j$, so every factor
is at most its value at $j=0$, and the second because
$n_1/n\geq f_1$ under (A3). The union bound over the $H$ cells then
gives $\Pr(\Omega_0^c\mid s)\leq H(1-f_1)^{n_{\min}}$ with
$n_{\min}=\min_hn_h$.

\emph{(ii) Expansion of $E_q(1/n_{1h})$.} Fix $h$, write $X=n_{1h}$
and $\mu=E_q(X)=n_1n_h/n$. For $X\geq 1$ the identity
\[
\frac{1}{X}=\frac{1}{\mu}+\frac{\mu-X}{\mu^2}
+\frac{(\mu-X)^2}{\mu^2X}
\]
holds exactly: multiplying the right-hand side by $\mu^2X$ gives
$\mu X+(\mu-X)X+(\mu-X)^2=\mu^2$. It is an algebraic identity, not a
Taylor expansion, and it is valid on $\Omega_0$, where $X\geq 1$.
Using $0\leq(\mu-X)^2/X\leq(\mu-X)^2$ for $X\geq 1$ and taking
conditional expectations,
\[
\Bigl|E_q\Bigl(\frac{1}{X}\,\Big|\;\Omega_0\Bigr)-\frac{1}{\mu}\Bigr|
\leq\frac{|E_q(\mu-X\mid\Omega_0)|}{\mu^2}
+\frac{E_q\{(\mu-X)^2\mid\Omega_0\}}{\mu^2}.
\]
Unconditionally $E_q(\mu-X)=0$ and, for the hypergeometric law,
$V_q(X)=\mu\bigl(1-n_h/n\bigr)\{(n-n_1)/(n-1)\}\leq\mu$.
Conditioning on $\Omega_0$ perturbs each moment by at most
$Cn\Pr(\Omega_0^c\mid s)/\Pr(\Omega_0\mid s)$, since $|\mu-X|\leq n$
and the conditional and unconditional expectations of a variable
bounded by $M$ differ by at most $M$ times the odds of the
complement of the conditioning event; the bounded variables here are
$\mu-X$ and $(\mu-X)^2$, while $1/X$ itself is only ever handled
conditionally on $\Omega_0$, which guarantees $X\geq 1$ and makes
the expansion legitimate. By the bound just proved this
perturbation is exponentially small in $n_{\min}$, and it is
absorbed into the constant when $f_1n_{\min}\geq 2$. Hence
$E_q(1/X\mid\Omega_0)=\mu^{-1}\{1+O(\mu^{-1})\}$.

Finally, with $\mu=n_1n_h/n$,
\[
\frac{1}{\mu}-\frac{1}{n_h}
=\frac{1}{n_h}\Bigl(\frac{n}{n_1}-1\Bigr)
=\frac{1}{n_h}\cdot\frac{n-n_1}{n_1}
=\frac{1-f_1}{f_1n_h}\bigl\{1+O(n^{-1})\bigr\},
\]
the $O(n^{-1})$ accounting for $n_1=\lceil f_1n\rceil$ in place of
$f_1n$, in the sense of \eqref{eq:rounding}. Collecting the two
error terms yields the displayed form of the theorem with
$|r_h|\leq C_2/(f_1n_h)$ for an absolute constant $C_2$. \qed

\subsection{Proof of Corollary~\ref{cor:relative-cost}}
\label{app:relative-cost}

We prove the result for the cell-stratified second phase of
Theorem~\ref{thm:trees}(i) with $n_{1h}=\lceil f_1n_h\rceil$; under
global $\mathrm{SRSWOR}$ the statement differs by the exponentially
small event $\Omega_0^c$ and by the remainders $r_h$ of
Theorem~\ref{thm:trees}(ii), neither of which affects the order. All
order symbols are uniform in $h$ under the stated cell-regularity
conditions, and $C$ denotes a generic positive constant that may
change between occurrences. The structure of the argument is: obtain
$V_1\asymp(N^2/n)\bar{S}^2$; obtain
$E_p(V_2)\asymp\{(1-f_1)/f_1\}(H/n)(N^2/n)\bar{S}^2$; divide. In
both steps the work consists of controlling the cells that are
atypically small, which is what the concentration inequalities do.

\emph{First phase.} Under $\mathrm{SRSWOR}(n)$ with the
cell-stratified second phase, $\bar{m}_s$ equals the full-sample cell
mean on each cell (the argument of Theorem~\ref{thm:trees}(ii) with
$f_1=1$), so, using $\pi_k=n/N$,
$\hat{Y}_{RB}=\sum_hN_h\bar{y}_{s_h}$: the classical
poststratified estimator. Indeed the correction term
$\sum_{k\in s}(y_k-\bar{m}_s(\mathbf{x}_k))/\pi_k$ vanishes because
each cell's residuals sum to zero over $s_h$, by the same
mechanism as in Corollary~\ref{cor:fullfit-plugin}.

Let $\mathbf{n}=(n_1^{\mathrm{c}},\ldots,n_H^{\mathrm{c}})$ denote
the first-phase cell counts, with $n_h^{\mathrm{c}}=n_h$.
Conditionally on $\mathbf{n}$ with all $n_h\geq 1$, the sets
$s_h$ are independent $\mathrm{SRSWOR}(n_h)$ samples within the
cells (the same factorization argument as in
Appendix~\ref{app:trees}), so by (P2)
\begin{equation}
\label{eq:v1-ps}
V_p(\hat{Y}_{RB}\mid\mathbf{n})
=\sum_{h}N_h^2\Bigl(\frac{1}{n_h}-\frac{1}{N_h}\Bigr)S^2_{y,U,h},
\end{equation}
and $E_p(\bar{y}_{s_h}\mid\mathbf{n})=\bar{y}_{U,h}$, free of
$\mathbf{n}$, makes the between-counts component of the variance
vanish.

Let $E_n^{(1)}=\{|n_h-nP_h|\leq nP_h/2\ \forall h\}$ with
$P_h=N_h/N$. Since $n_h$ is a sum of $n$ without-replacement draws of
the bounded indicators $\mathbf{1}(k\in A_h)$, the inequalities of
\citet[Sec.~6]{hoeffding1963} and \citet{serfling1974} give
$\Pr\{(E_n^{(1)})^c\}\leq 2H\exp(-Cn/H)\to 0$ when $n/H\to\infty$.
On $E_n^{(1)}$ the cell-regularity conditions give $n_h\asymp n/H$
and $1/n_h-1/N_h\asymp(1-\kappa)H/n$ since $n/N\to\kappa<1$, so,
using $N_h\asymp N/H$ and
$S^2_{y,U,h}\asymp\bar{S}^2$, the sum \eqref{eq:v1-ps} is bounded
above and below by constants times $(N^2/n)\bar{S}^2$. On the
complement, \eqref{eq:v1-ps} is at most $CN^2\bar{S}^2$ crudely (each
term being bounded by $N_h^2S^2_{y,U,h}$), so its contribution to
$V_1=E_p\{V_p(\hat{Y}_{RB}\mid\mathbf{n})\}$ is
$O(N^2He^{-Cn/H})=o(N^2/n)$. Hence $V_1\asymp(N^2/n)\bar{S}^2$.

\emph{Second phase.} By Theorem~\ref{thm:trees}(i) with
$n_{1h}=\lceil f_1n_h\rceil$,
\[
V_2=\sum_hD_h^2\Bigl(\frac{1}{n_{1h}}-\frac{1}{n_h}\Bigr)S^2_{y,s,h},
\qquad
\frac{1}{n_{1h}}-\frac{1}{n_h}
\asymp\frac{1-f_1}{f_1}\cdot\frac{1}{n_h},
\]
the equivalence requiring $f_1n_h\to\infty$, which is the regime
condition $f_1n/H\to\infty$ combined with $n_h\asymp n/H$; this is
the precise point at which that condition is used, and the reason
\eqref{eq:relative-cost} overstates the second-phase share when it
fails.

Under $\mathrm{SRSWOR}(n)$ we have $D_h=N_h-(N/n)n_h$, a centred
multiple of the cell count, so
\[
E_p(D_h^2)=\Bigl(\frac{N}{n}\Bigr)^2V_p(n_h)
=\Bigl(\frac{N}{n}\Bigr)^2nP_h(1-P_h)\frac{N-n}{N-1}
\;\asymp\;\frac{N^2}{nH},
\]
using $P_h\asymp 1/H\to 0$ and $\kappa<1$. For the upper bound,
$S^2_{y,s,h}\leq C$ by (A2), and on $E_n^{(1)}$
$1/n_{1h}-1/n_h\leq(1-f_1)/(f_1n_{1h}-f_1)\leq C(1-f_1)H/(f_1n)$, so
\[
E_p\bigl(V_2\mathbf{1}_{E_n^{(1)}}\bigr)
\leq C\,\frac{1-f_1}{f_1}\cdot\frac{H}{n}\sum_hE_p(D_h^2)
\leq C\,\frac{1-f_1}{f_1}\cdot\frac{H}{n}\cdot\frac{N^2}{n},
\]
since the $H$ terms $E_p(D_h^2)\asymp N^2/(nH)$ sum to $N^2/n$. On
the complement $V_2\leq CN^2H(H/n)$ crudely, and the contribution is
again exponentially negligible.

For the lower bound, let
$E_n^{(2)}=E_n^{(1)}\cap\{S^2_{y,s,h}\geq S^2_{y,U,h}/2\ \forall h\}$.
The sample variance within a cell is a smooth function of two
without-replacement means of bounded variables, so the same
concentration inequalities give
$\Pr\{(E_n^{(2)})^c\}\leq CH\exp(-Cn/H)$. Then
\[
E_p(V_2)
\geq C\,\frac{1-f_1}{f_1}\cdot\frac{H}{n}
\sum_hE_p\bigl(D_h^2\mathbf{1}_{E_n^{(2)}}\bigr)
\geq C\,\frac{1-f_1}{f_1}\cdot\frac{H}{n}
\sum_h\Bigl\{E_p(D_h^2)
-\bigl[E_p(D_h^4)\Pr\{(E_n^{(2)})^c\}\bigr]^{1/2}\Bigr\},
\]
by Cauchy--Schwarz applied to
$E_p(D_h^2\mathbf{1}_{(E_n^{(2)})^c})$. The fourth central moment of
a hypergeometric count satisfies
$E_p\{(n_h-nP_h)^4\}\leq C\{(nP_h)^2+nP_h\}$, whence
$E_p(D_h^4)=(N/n)^4E_p\{(n_h-nP_h)^4\}
\leq C\{N^2/(nH)\}^2(1+H/n)$, so the subtracted term is
$o\{E_p(D_h^2)\}$ because $H\exp(-Cn/H)\to 0$ when $n/H\to\infty$.

Combining the two bounds,
$E_p(V_2)\asymp\{(1-f_1)/f_1\}(H/n)(N^2/n)\bar{S}^2$, the
comparability of the $S^2_{y,U,h}$ across cells tying the constants
to $\bar{S}^2$. Dividing by $V_1\asymp(N^2/n)\bar{S}^2$, the factors
$N^2/n$ and $\bar{S}^2$ cancel and \eqref{eq:relative-cost}
follows. \qed

\subsection{Proof of Proposition~\ref{prop:v2-known}}
\label{app:v2-known}

Under the cell-stratified design the training cell sizes $n_{1h}$ are
fixed by the analyst, and the remaining ingredients of
\eqref{eq:v2hat} --- the frame counts $N_h$, the weighted counts
$\hat{N}_{\pi,h}$, the cell sample sizes $n_h$ and the within-cell
sample variances $S^2_{y,s,h}$ --- are functions of $s$ and the
frame. Hence $\hat{V}_2$ is $s$-measurable, and it coincides term by
term with the exact expression \eqref{eq:v2-trees-exact}: the two
displays are literally the same formula, so $\hat{V}_2=V_2$ with no
expectation taken. This is the sense in which the second-phase
variance is known rather than estimated; the reason it is available
is the structural reversal of Section~\ref{sec:setup-classical},
namely that $y$ is observed on all of $s$, so $S^2_{y,s,h}$ is
computable.

Under global $\mathrm{SRSWOR}(n_1)$ the counts $n_{1h}$ are random
and $\hat{V}_2$ is no longer $s$-measurable, but the other factors
still are, so by (P3), on $\Omega_0$,
\[
E_q(\hat{V}_2\mid s,\Omega_0)
=\sum_hD_h^2\Bigl\{E_q\Bigl(\frac{1}{n_{1h}}\,\Big|\;\Omega_0\Bigr)
-\frac{1}{n_h}\Bigr\}S^2_{y,s,h}
=V_q(T_1\mid s,\Omega_0),
\]
the last equality being \eqref{eq:v2-trees-global}. Thus $\hat{V}_2$
is exactly conditionally unbiased in that case. \qed

\subsection{Proof of Proposition~\ref{prop:rep-unbiased}}
\label{app:rep-unbiased}

Condition on $s$ throughout. The replicate subsamples
$s_1^{*(1)},\ldots,s_1^{*(A)}$ are drawn independently from the same
mechanism $q(\cdot\mid s)$ that generated the realized $s_1$, so the
statistics $T_1^{*(a)}
=\sum_{k\in U}(1-I_k/\pi_k)\hat{m}_{s_1^{*(a)}}(\mathbf{x}_k)$,
$a=1,\ldots,A$, are conditionally independent and identically
distributed, with common conditional mean
$\theta:=E_q(T_1'\mid s)$ and common conditional variance
$V_q(T_1'\mid s)=V_2$ --- the same $V_2$ as in
Theorem~\ref{thm:decomp}, since $T_1$ and $T_1'$ differ by an
$s$-measurable term and (P3) applies.

That the sample variance of independent and identically distributed
variables is unbiased for their common variance is standard; we
give the computation because it is what makes the statement exact
for every $A\geq 2$ and every algorithm. With
$\bar{T}^*_1=A^{-1}\sum_aT_1^{*(a)}$, the algebraic identity
\[
\sum_{a=1}^{A}\bigl(T_1^{*(a)}-\bar{T}^*_1\bigr)^2
=\sum_{a=1}^{A}\bigl(T_1^{*(a)}-\theta\bigr)^2
-A\bigl(\bar{T}^*_1-\theta\bigr)^2
\]
holds for any $\theta$. Taking $E_q(\cdot\mid s)$, the first sum has
expectation $AV_2$, while by independence
$E_q\{(\bar{T}^*_1-\theta)^2\mid s\}=V_2/A$, so the right-hand side
has expectation $AV_2-V_2=(A-1)V_2$. Dividing by $A-1$ gives
$E_q(\hat{V}_2^{\mathrm{rep}}\mid s)=V_2$ exactly.

No property of the algorithm has been used: $\mathcal{A}$ may be a
forest, a boosted ensemble, or any procedure with internal
randomization, provided only that the replicate draws reproduce the
same randomization as the realized one --- including that internal
randomness, when the reported point estimator uses it. Finally,
taking $E_p$ and using $E=E_pE_q$,
$E(\hat{V}_2^{\mathrm{rep}})
=E_p\{E_q(\hat{V}_2^{\mathrm{rep}}\mid s)\}=E_p(V_2)$, the
second-phase component of \eqref{eq:main-decomp}. \qed


\end{document}


\maketitle

\section{A variant restricting the correction to the training
sample}
\label{supp:variant}

This section examines the variant mentioned in the Remark on the
choice of the residual-correction sample in the main text: the
construction suggested by a literal transcription of the classical
two-phase regression estimator, which corrects only over $s_1$,
with two-phase weights $\pi^*_k=\pi_k\,n_1/n$,
\[
\hat{Y}_{\mathrm{alt}}
=\sum_{k\in U}\hat{m}_1(\mathbf{x}_k)
+\sum_{k\in s_1}\frac{y_k-\hat{m}_1(\mathbf{x}_k)}{\pi^*_k}.
\]

In the degenerate case ($\hat{m}_1\equiv\bar{y}_{s_1}$), under
$\mathrm{SRSWOR}$ at both phases, the two-phase weights are
$\pi^*_k=(n/N)(n_1/n)=n_1/N$, so $1/\pi^*_k=N/n_1$ and
\[
\hat{Y}_{\mathrm{alt}}
=\sum_{k\in U}\bar{y}_{s_1}
+\sum_{k\in s_1}\frac{y_k-\bar{y}_{s_1}}{\pi^*_k}
=N\bar{y}_{s_1}
+\frac{N}{n_1}\sum_{k\in s_1}(y_k-\bar{y}_{s_1})
=N\bar{y}_{s_1},
\]
because the residuals of a sample mean sum to zero over the very
set on which the mean was computed. The correction term is
therefore identically zero: the values $y_k$ observed on $s_2$
never enter, and the estimator uses only the $n_1<n$ observations
in $s_1$.

For the efficiency comparison, note that marginally $s_1$ is an
$\mathrm{SRSWOR}(n_1)$ sample \emph{from $U$} (fact~(P4) of the
main text's Appendix): drawing $n$ units at random from $U$ and
then $n_1$ of them at random makes every subset of $U$ of size
$n_1$ equally likely. Hence the textbook mean-of-an-SRSWOR-sample
formula applies with $D=U$ to both estimators, and
\[
V(N\bar{y}_s)=N^2\Bigl(\frac{1}{n}-\frac{1}{N}\Bigr)S^2_{y,U}
\;<\;
N^2\Bigl(\frac{1}{n_1}-\frac{1}{N}\Bigr)S^2_{y,U}
=V(N\bar{y}_{s_1})
\]
for $n_1<n$, both estimators being exactly unbiased in this case.
The variance ratio is $(1/n_1-1/N)/(1/n-1/N)$, which for
$n/N\to 0$ tends to $1/f_1$: at $f_1=1/2$ the variant doubles the
variance.

The comparison isolates, in the cleanest case, why the main text
pools the residual correction over all of $s$: restricting the
correction to $s_1$ discards the assessment sample entirely, and
with it the very observations whose out-of-sample character
motivates the construction. The exactly unbiased single-split
estimator of Sanguiao Sande and Zhang (2021) restricts the
correction to $s_2$ instead, with conditional weights; its exact
relation to the pooled form is the composite representation of the
main text.

\section{Reproducibility}

The R scripts distributed with the paper regenerate every table
and figure, and verify every exact identity of the main text ---
including the composite representation and the vanishing of the
full-fit correction term --- by Monte Carlo simulation or, where
the design is small enough, by complete enumeration.